\documentclass[aps,prb,reprint,floatfix,superscriptaddress,longbibliography]{revtex4-2}
\usepackage{amsmath,amssymb,bm,graphicx,booktabs}
\usepackage{placeins}
\usepackage[colorlinks=true,allcolors=blue]{hyperref}

\newcommand{\Tr}{\operatorname{Tr}}

\newcommand{\PaperTitle}{Chiral symmetry and the orbital susceptibility of disordered honeycomb flakes}

\begin{document}
\title{\PaperTitle}
\author{B. Murzaliev}
\affiliation{Materials \& Structural Analysis, Thermo Fisher Scientific, De Schakel 2, 5651 GH Eindhoven, The Netherlands}

\begin{abstract}
The orbital susceptibility of a small disordered grain is dominated by rare samples in which an occupied and an empty level nearly touch.
In a chaotic grain such near-degeneracies are suppressed by level repulsion and must also coincide with the chemical potential, so the mean susceptibility stays finite.
In a neutral honeycomb flake, chiral (sublattice) symmetry removes the second condition: It places the chemical potential in the middle of the smallest gap, in every disorder realization.
Here we show that the mean orbital susceptibility then grows as $\ln(1/T)$ at low temperature $T$ if disorder can close this gap while a magnetic field still mixes the two levels, and that the hopping geometry decides whether it can: Positive nearest-neighbor hopping on a compact hexagonal flake keeps the gap open, third-neighbor hopping allows it to close.
In a chiral random-matrix model with $n$ level pairs the coefficient of the logarithm is $C_n=\sqrt{2/n}\,\Gamma[(n+1)/2]/\Gamma(n/2)$, close to unity, and a detuned chemical potential cuts the logarithm off.
\end{abstract}
\maketitle
\raggedbottom

The orbital magnetic response of graphene reflects its Dirac spectrum.\cite{castroNeto2009}
At charge neutrality, the clean bulk has a singular diamagnetism, predicted by McClure\cite{mcclure1956,fukuyama2007} and observed in a graphene monolayer.\cite{vallejo2021}
In a finite flake whose level spacing exceeds the temperature, the response becomes particularly sensitive to the few levels closest to the chemical potential.
Its sign and magnitude depend on the boundary, filling, and wave functions.\cite{ominato2012,ominato2013,hesse2014,raoux2014}
Disorder makes these levels, and with them the magnetic response, random from flake to flake.

A nearly degenerate pair of occupied and empty levels gives a large paramagnetic response: magnetic mixing lowers the energy of the occupied state by an amount inversely proportional to the level separation.\cite{serota2001}
Rare small gaps therefore produce broad distributions of orbital susceptibility.\cite{serota2002,goldstein2004}
In a chaotic grain, level repulsion suppresses small gaps, and a fixed chemical potential must also fall inside the gap, so the ensemble mean remains finite.\cite{goldstein2004}
In integrable rectangular grains, whose levels do not repel, the mean inverse gap diverges logarithmically.\cite{lou2007}
A graphene flake offers a third possibility.
Its chiral (sublattice) symmetry pairs the energies at $\pm s$, with $s>0$, and places the neutral chemical potential at the center of the smallest gap in every realization: No coincidence is needed.
The nonzero levels still repel one another, but the innermost pair can approach zero.

\begin{figure}[tbp]
\centering
\includegraphics[width=\columnwidth]{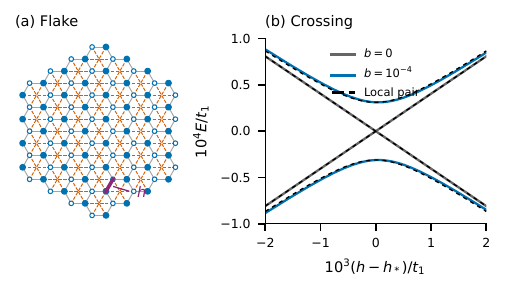}
\caption{A zero-energy pair and its magnetic splitting.
(a) A 96-site flake with first-neighbor (gray) and third-neighbor (orange) bonds; filled and open sites mark the two sublattices.
(b) Varying the hopping $h$ of the purple bond through its crossing value $h_*/t_1\simeq1.084$ closes the central gap at zero field (gray).
A field $b=10^{-4}$ opens an avoided crossing (blue).
Solid curves show the single-particle energies $E$ near zero; dashed curves show Eq.~(\ref{eq:local}).
Here $t_1$ and $t_3=0.2t_1$ are the first- and third-neighbor hopping scales; all bond factors lie in $(0.7,1.3)$.
The dimensionless field is $b=BL_0^2/\Phi_0$, with magnetic field $B$, nearest-neighbor distance $L_0$, and single-electron flux quantum $\Phi_0$.}
\label{fig:crossing}
\end{figure}

Here we show that the hopping geometry decides whether this route is available in a finite honeycomb flake.
For nearest-neighbor hopping amplitudes drawn from a positive interval bounded away from zero, a compact hexagonal flake has a nonzero minimum gap, and its mean susceptibility saturates upon cooling.
Third-neighbor bonds preserve chiral symmetry\cite{sticlet2013} and allow a zero-energy crossing that couples to the field [Fig.~\ref{fig:crossing}].
The neighborhoods of such crossings give a divergent ensemble mean even though almost every individual flake has a finite zero-temperature response.
We relate the coefficient of the logarithm to the density of such crossings weighted by their magnetic coupling, compare it with the full lattice response, and obtain an exact counterpart in a chiral random-matrix ensemble.

\textit{Pair mechanism.}
We consider spinless, noninteracting electrons on a bipartite lattice whose two sublattices have $n+\nu$ and $n$ sites; the integer $\nu\geq0$ measures their imbalance.
Hopping between opposite sublattices gives the Hamiltonian
\begin{equation}
H(q)=\begin{pmatrix}0&W(q)\\W^\dagger(q)&0\end{pmatrix}.
\label{eq:chiral}
\end{equation}
The $(n+\nu)\times n$ matrix $W$ contains the hopping amplitudes; $W^\dagger$ is its Hermitian conjugate.
The operator that reverses the sign of the wave function on one sublattice anticommutes with $H$; this is the chiral symmetry.
For full-rank $W$, the spectrum contains $n$ pairs $\pm s_i$, where $s_i>0$ are its singular values ($i=1,\ldots,n$), and $\nu$ protected zero modes.
The parameter $q$ breaks time-reversal symmetry: it denotes the dimensionless magnetic field $b$ on the lattice or the Gaussian crossover parameter $a$ below.
Reversing the field gives $H(-q)=H(q)^*$, so $H(0)$ is real.

For temperature $T$ and chemical potential $\mu$, the grand potential is $\Omega=-T\Tr\ln[1+e^{-(H-\mu)/T}]$, with Boltzmann's constant $k_B=1$ and the trace over single-particle states.
We define the response $\chi_q(T,q;\mu)=-\partial_q^2\Omega$ at fixed $T$ and $\mu$, positive for paramagnetism, and suppress $\mu$ at neutrality.
We take the zero-field derivative at finite $T$ before lowering the temperature, keeping the size fixed.
For the lattice, the pair decomposition uses the symmetric Peierls gauge.
At $\mu=q=0$, magnetic mixing between the normalized partner states $|\pm i\rangle$ contributes
\begin{equation}
\chi_i^{\rm pair}(T)=\frac{|J_i|^2}{s_i}\tanh\frac{s_i}{2T},
\qquad J_i=\langle+i|\partial_qH|-i\rangle.
\label{eq:pair}
\end{equation}
Here $J_i$ is the mixing matrix element per unit $q$, evaluated at zero field.
For $s_i\gg T$, only the lower level is occupied, giving the inverse-gap response $|J_i|^2/s_i$.
For $s_i\ll T$, thermal occupation of both partners limits the response to $|J_i|^2/(2T)$.

Averaging over disorder gives
\begin{equation}
\begin{aligned}
\Big\langle\sum_i\chi_i^{\rm pair}\Big\rangle
 &=\int_0^\infty ds\,\mathcal A(s)\frac{\tanh(s/2T)}s,\\
\mathcal A(s)&=\Big\langle\sum_i|J_i|^2\delta(s-s_i)\Big\rangle.
\end{aligned}
\label{eq:weighted}
\end{equation}
The weighted density $\mathcal A(s)$ combines the probability of a small pair energy with its magnetic coupling; $\delta$ is the Dirac delta function and brackets denote the disorder average.
If $\mathcal A(s)$ tends to a positive constant at zero energy, every decade of pair energies above $T$ contributes equally, and the sum grows as $\mathcal A(0)\ln(1/T)$.
The full response also contains the mixing with all other levels and the explicit field curvature of the Hamiltonian, both retained in our calculations.\cite{goldstein2004,supplemental}

\textit{Honeycomb flakes.}
We take compact hexagonal flakes, which contain every hexagon inside their boundary, with $N=2n$ sites, equal sublattice populations ($\nu=0$), and open zigzag edges.
Independent factors uniform in $[0.7,1.3]$ multiply the first- and third-neighbor hopping scales $t_1$ and $t_3$.
A perpendicular field $B$ enters through Peierls phases in the symmetric gauge.
We write $b=BL_0^2/\Phi_0$, where $L_0$ is the nearest-neighbor distance and $\Phi_0=2\pi\hbar/|e|$ the single-electron flux quantum.
We measure energies, temperatures, and $\chi_b$ in units of $t_1$; the physical susceptibility per spinless flake is $t_1(L_0^2/\Phi_0)^2\chi_b$.

With nearest-neighbor hopping alone, the determinant of $W$ is a sum over dimer coverings, sets of bonds that touch every site exactly once.\cite{kenyon2009}
On a compact flake all coverings enter with the same sign,\cite{supplemental} so for positive hopping the determinant cannot vanish.
Since the hopping amplitudes are bounded away from zero, the gap then has a positive minimum at each size, and the mean susceptibility saturates as $T\to0$.
The argument relies on the compact shape and on the lower bound of the hopping; the thermodynamic limit of random-hopping systems is a separate question.\cite{gade1993,motrunich2002,nayak2024}

Third-neighbor bonds add determinant terms of opposite sign, so the gap can close.
For $N=96$ and $t_3/t_1=0.2$, we construct an exact crossing with all bond factors inside the same interval.\cite{supplemental}
Varying one hopping amplitude $h$ through its crossing value $h_*$ gives the local energies
\begin{equation}
\frac{E_\pm}{t_1}\simeq\pm\sqrt{g^2[(h-h_*)/t_1]^2+j_b^2b^2}.
\label{eq:local}
\end{equation}
The dimensionless slope $g=0.040$ measures how fast the gap opens with $h$, and $j_b=-0.31$ is the magnetic mixing amplitude in units of $t_1$ per unit $b$; both are evaluated at $h=h_*$ and $b=0$.
Small changes in other bonds shift the crossing without removing it, producing a continuous family of crossings that couple to the field.

Integrating the pair term, Eq.~(\ref{eq:pair}), across these crossings bounds the mean susceptibility from below by a positive multiple of $\ln(1/T)$, minus a constant.\cite{supplemental}
The other mixing terms are nonnegative and the direct field curvature is bounded, so they cannot remove this divergence.
Summed over all crossings, the density $\mathcal A(0)$ of Eq.~(\ref{eq:weighted}) is finite.\cite{supplemental}
Each crossing enters with its squared magnetic coupling divided by the rate at which the gap opens as the bonds change, because a slowly opening gap occupies more disorder volume per energy interval.

\begin{figure}[tbp]
\centering
\includegraphics[width=\columnwidth]{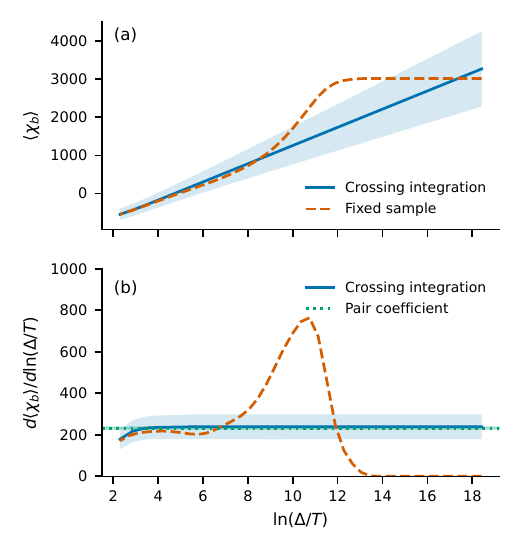}
\caption{Rare pairs control the low-temperature mean response.
(a) Mean susceptibility of neutral 96-site flakes at $t_3/t_1=0.4$, including all levels.
Blue: average obtained by integrating one bond continuously across the small-gap regions.
Orange dashes: a fixed sample of 20,007 flakes, which misses the smallest gaps and levels off.
(b) Logarithmic slope of the blue curve, compared with the coefficient $\mathcal A(0)=231.5$ obtained independently from the crossings (green).
Bands: $1.96$ standard errors.
Here $T$ is temperature, $\Delta$ the mean level spacing near zero energy, and energies are in units of the nearest-neighbor hopping $t_1$.}
\label{fig:thermal}
\end{figure}

\textit{Disorder-averaged response.}
Figure~\ref{fig:thermal} compares the temperature dependence of the full mean response with the coefficient obtained independently from the crossings, for $N=96$ and $t_3/t_1=0.4$.
Here $\Delta$ is the mean level spacing near zero energy, the ensemble average of the spacings between the third and tenth positive levels.
A fixed collection of flakes has a smallest sampled gap and therefore appears to saturate below that scale.
We recover the missing rare events by integrating one bond continuously through the crossing region before averaging over the other bonds.\cite{supplemental}
This calculation reaches $T/\Delta=10^{-8}$, where the full response continues to rise with $\ln(\Delta/T)$.
Its slope, $239\pm31$, agrees with $\mathcal A(0)=231.5\pm2.2$ obtained from the crossings (standard errors), so the crossings fix the coefficient of the logarithm of the full response.

At $T=0.003\Delta$ about two thirds of the flakes are diamagnetic, and the paramagnetic mean comes from the few with a nearly closed pair;\cite{supplemental} a mean over a finite sample therefore depends on which of these rare flakes the sample happens to contain.
At $t_3/t_1=0.2$ the exact crossing proves the divergence although random sampling finds no crossings, whereas at $t_3/t_1=0.4$ the crossings are frequent enough to resolve numerically.
Stronger bond disorder makes small gaps more frequent and lowers the value of $t_3$ at which they appear; the coefficient of the logarithm remains $\mathcal A(0)$.\cite{supplemental}
At $t_3/t_1=0.4$ the coefficient grows with size, $\mathcal A(0)=232$, 804, and 1353 (in units of $t_1$) for $N=96$, 216, and 384, but not in proportion to the area.\cite{supplemental}

\begin{figure}[tbp]
\centering
\includegraphics[width=0.94\columnwidth]{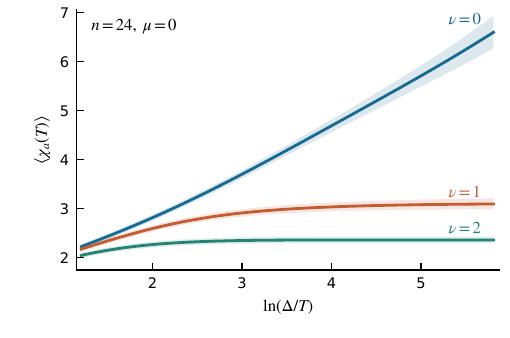}
\caption{Sublattice imbalance removes the logarithm of the random-matrix model.
The full mean response $\langle\chi_a(T,0)\rangle$ grows for $\nu=0$ and saturates for $\nu=1,2$, where $\nu$ counts protected zero modes.
All curves use $n=24$ positive levels, $\mu=0$, Gaussian energy unit $E_0=1$, and reference spacing $\Delta=\pi/(2n)$.
Each curve averages 6,000 independent matrices, with the average over $Y$ taken analytically; bands give pointwise 95\% bootstrap intervals.}
\label{fig:gaussian}
\end{figure}

\textit{Random-matrix model.}
The pair mechanism becomes exactly solvable when $W$ has independent Gaussian entries, as in the chiral random-matrix ensembles.\cite{verbaarschot1994,altland1997}
We choose $W(a)=(X+iaY)/\sqrt{1+a^2}$,\cite{nishigaki2012b} where $X,Y$ are independent real $(n+\nu)\times n$ matrices with zero-mean entries of variance $1/n$, so that $W$ is real (chiral orthogonal class) at $a=0$ and complex (chiral unitary class) at $a=1$.
Energies are measured in the unit $E_0$ of the matrix entries, which we set to 1; the normalization keeps the mean squared hopping constant as $a$ breaks time-reversal symmetry.
The conversion of $a$ to a physical flux depends on the system.\cite{goldstein2004,murzaliev2019}

Averaging first over $Y$ and then integrating by parts in the singular values of $X$ expresses the full response through the positive-level density:\cite{supplemental}
\begin{equation}
\langle\chi_a(T,0;\mu)\rangle=\frac1n\int_0^\infty ds\,R_{n,\nu}(s)
\left[\frac{q_\mu(s)}s-q_\mu'(s)\right].
\label{eq:identity}
\end{equation}
Here $R_{n,\nu}(s)=\langle\sum_i\delta(s-s_i)\rangle$, normalized to $n$, and $q_\mu(s)=f(-s)-f(s)$ is the occupation difference, with $f(E)=[1+e^{(E-\mu)/T}]^{-1}$.
The prime denotes differentiation with respect to $s$.
Equation~(\ref{eq:identity}) includes the direct curvature of $W(a)$ and mixing among all levels.
For balanced sublattices, the Laguerre--Selberg integral gives $R_{n,0}(0)=nC_n$, yielding\cite{selberg2008,wirtz2015}
\begin{equation}
\begin{aligned}
\langle\chi_a(T,0)\rangle&=C_n\ln(\Delta/T)+O_n(1),\\
C_n&=\sqrt{\frac2n}\frac{\Gamma[(n+1)/2]}{\Gamma(n/2)}.
\end{aligned}
\label{eq:gaussian-log}
\end{equation}
Here $\Gamma$ is the gamma function, $\Delta=\pi/(2n)$ is the level spacing near zero energy obtained from the macroscopic density, and $O_n(1)$ stays bounded as $T\to0$ at fixed $n$; $C_n$ approaches unity for large $n$.
Averaging over $Y$ at fixed $X$ gives $\overline{|J_i|^2}=1/n$, so $\mathcal A(0)=C_n$: the logarithm comes from the mixing of the two partners, and the second term of Eq.~(\ref{eq:identity}) adds only a finite correction.

The density at zero energy stays finite even though the positive levels repel one another.
Their joint density contains the factors $s_i^\nu$ and $|s_i^2-s_j^2|$, so for $\nu=0$ a level near zero is repelled only by the other levels, which leaves a finite weight, and $\det X$ changes sign across the surface where that level vanishes, as at the lattice crossing.
Positive nearest-neighbor hopping on a compact flake excludes this surface.

Equation~(\ref{eq:identity}) also shows how the logarithm is removed.
A chemical potential $\mu\neq0$ makes partners with $s<|\mu|$ both occupied or both empty, which cuts off the growth below $T\sim|\mu|$.
A sublattice imbalance $\nu>0$ gives $R_{n,\nu}(s)\propto s^\nu$ near zero, because the protected zero modes repel the nonzero levels, and the inverse-gap integral converges; the zero modes themselves stay pinned and contribute no field curvature.
Figure~\ref{fig:gaussian} shows both cases for $n=24$ at $\mu=0$: the balanced response grows with slope $C_{24}\simeq0.99$, and the imbalanced ones saturate.
As a check, estimates from the level curvatures and from Eq.~(\ref{eq:identity}), on independent samples, agree within their statistical errors.\cite{supplemental}

Along the crossover the mean response changes sign.
The symmetry $a\to1/a$ of $W(a)$ makes the response at finite $a$ integrate to zero over $0<a<1$, and since it is positive at $a=0$ it must be negative in between.\cite{supplemental}
A similar change from paramagnetic to diamagnetic response along a crossover to the unitary ensemble was found for metallic grains.\cite{murzaliev2019}

\begin{figure}[tbp]
\centering
\includegraphics[width=0.95\columnwidth]{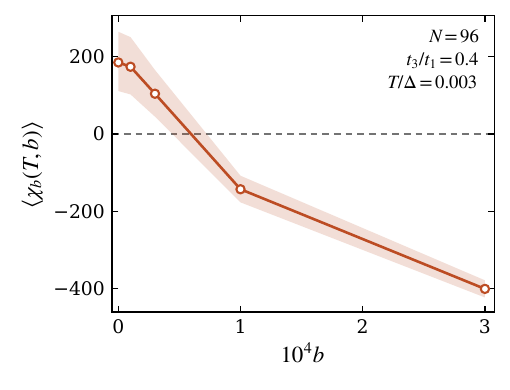}
\caption{A magnetic field reverses the sign of the response.
Mean differential susceptibility $\chi_b=-\partial_b^2\Omega$, evaluated at the field $b$ and including all levels, for 20,007 neutral 96-site flakes with $t_3/t_1=0.4$ at $T=0.003\Delta$; the same flakes enter every field.
Band: pointwise 95\% bootstrap interval.
Here $b=BL_0^2/\Phi_0$, with magnetic field $B$, nearest-neighbor distance $L_0$, and flux quantum $\Phi_0$; for $L_0=0.142$~nm, $b=10^{-4}$ corresponds to 20.5~T. Energies in units of $t_1$.}
\label{fig:field}
\end{figure}

\textit{Finite field.}
A magnetic field opens the avoided crossing of Eq.~(\ref{eq:local}) and suppresses its large low-field curvature once the magnetic splitting exceeds the zero-field gap or the temperature.
The pair alone still gives a nonnegative differential response at any field,\cite{supplemental} but the mixing with other levels and the direct field curvature can now dominate, and the mean differential susceptibility turns diamagnetic between $b=3\times10^{-5}$ and $10^{-4}$ (Fig.~\ref{fig:field}).
A negative differential response means that the magnetic moment decreases with field, whatever its sign.

\textit{Conclusion.}---In summary, we have shown that the mean orbital susceptibility of a neutral, disordered chiral flake grows as $\mathcal A(0)\ln(1/T)$, with $\mathcal A(0)$ the density of pair energies at zero weighted by the squared magnetic coupling of the pair.
Chiral symmetry places the chemical potential inside the smallest gap; the hopping geometry decides whether this gap can close.
Positive nearest-neighbor hopping on a compact flake gives $\mathcal A(0)=0$, third-neighbor hopping gives $\mathcal A(0)>0$, and the chiral random-matrix model gives $\mathcal A(0)=C_n$.
The key difference with the logarithm of integrable grains\cite{lou2007} is that no accidental coincidence of a small gap with the chemical potential is needed.

Our flakes idealize graphene in three respects.
A graphene parametrization gives $|t_3/t_1|\simeq0.095$,\cite{jung2013} about a third of the value at which near-zero levels become frequent in our ensembles, together with a second-neighbor hopping $|t_2/t_1|\simeq0.077$ that breaks chiral symmetry.
On-site disorder shifts small pairs away from the chemical potential: Site energies uniform in $[-v,v]$ with $v=10^{-2}t_1$ already turn the mean response negative at $T=0.003\Delta$.\cite{supplemental}
And on-site repulsion acts most strongly on the nearly closed pairs, whose two partner orbitals then carry nearly free spins; projecting a Hubbard interaction of strength comparable to $t_1$ onto these orbitals, we estimate that it cuts off the logarithm near $\Delta$.\cite{supplemental}

In closing, we comment on the experimental observability.
Magnetization measurements on graphene nanocrystals find a strong diamagnetism and, below 50~K, about one paramagnetic moment per crystallite.\cite{sepioni2010}
For $t_1=2.7\,\mathrm{eV}$ and $L_0=0.142\,\mathrm{nm}$, the 96-site flakes with $t_3/t_1=0.4$ have $\Delta\simeq86\,\mathrm{meV}$, so that $T=0.003\Delta$ is about $3\,\mathrm K$ and $b=10^{-4}$ corresponds to $20.5\,\mathrm T$.
The coefficient of the logarithm, $\mathcal A(0)\,t_1(L_0^2/\Phi_0)^2$, is $2.38\times10^{-27}\,\mathrm{J/T^2}$ per spinless flake, about 900 times smaller than the susceptibility of one free spin $1/2$ at $3\,\mathrm K$.
Prospects for detecting the logarithm in carbon flakes are therefore not promising.
A cleaner test is a four-site atomic loop, for which tunnel-coupled fermionic tweezer plaquettes\cite{yan2022} and synthetic fluxes for fermions\cite{mancini2015} provide the separate ingredients.
A square is the smallest loop whose determinant terms can cancel: For positive successive bond amplitudes $u_1,\ldots,u_4$, the condition $u_1u_3=u_2u_4$ closes the central gap, and a loop flux opens it.
Repeated draws of the bond amplitudes across this condition would reveal the logarithm in the current response to the synthetic flux.
With two spin-polarized atoms at fixed particle number the inverse-gap singularity remains,\cite{supplemental} and the Pauli principle removes the on-site interaction that limits the effect in carbon flakes.

From a more general perspective, the logarithm requires only a pair of levels that closes with a finite density while a perturbation still mixes it.
A short diffusive Josephson contact at phase difference $\phi=\pi$ realizes this situation---but now for Andreev levels.
A channel of transmission $\tau$ has levels at $\pm\Delta_S(1-\tau)^{1/2}$, with $\Delta_S$ the superconducting gap, the deviation $\phi-\pi$ mixes them as the field does here, and Dorokhov's distribution of transmissions\cite{dorokhov1984} gives a finite density of such pairs at zero energy.
The result is the logarithmic singularity at $\phi=\pi$ of the inverse inductance obtained by Kos, Nigg, and Glazman.\cite{kos2013}

\begin{acknowledgments}
I have benefited from discussions with M. I. Katsnelson and M. Titov.
\end{acknowledgments}

\paragraph*{Data availability.}
The ancillary files contain the exact crossing configuration and a script that verifies it in exact arithmetic; the data and code behind the numerical results are available from the author upon reasonable request.

\clearpage
%

\clearpage
\onecolumngrid
\setcounter{section}{0}
\setcounter{subsection}{0}
\setcounter{equation}{0}
\setcounter{figure}{0}
\setcounter{table}{0}
\renewcommand{\thefigure}{S\arabic{figure}}
\renewcommand{\theequation}{S\arabic{equation}}
\renewcommand{\thetable}{S\Roman{table}}
\renewcommand{\thesection}{S\Roman{section}}
\renewcommand{\theHsection}{supp.\arabic{section}}
\renewcommand{\theHsubsection}{supp.\arabic{section}.\arabic{subsection}}
\renewcommand{\theHequation}{supp.\arabic{equation}}
\renewcommand{\theHfigure}{supp.\arabic{figure}}
\renewcommand{\theHtable}{supp.\arabic{table}}
\setlength{\emergencystretch}{3em}
\phantomsection
\label{sm:start}
\pdfbookmark[0]{Supplemental Material}{supplement-title}
\begin{center}
\textbf{\large Supplemental Material}\\[0.5em]
\textbf{\PaperTitle}
\end{center}

This Supplemental Material contains the derivations and numerical details
behind the main text, in the order in which they appear there.
Section~\ref{sm:sec:response} gives the exact expression for the orbital
response and the numerical method. Section~\ref{sm:sec:lattice} treats the
honeycomb flakes: the obstruction for nearest-neighbor hopping, the
ensembles, and the exact crossing. Section~\ref{sm:sec:global-theory} proves
that the coefficient $\mathcal A(0)$ of the logarithm is finite, evaluates it,
and compares it with the full response. Section~\ref{sm:sec:gaussian}
derives the response of the random-matrix model and Sec.~\ref{sm:sec:finite-field}
the response at finite field. Sections~\ref{sm:sec:robustness} and
\ref{sm:sec:interaction} treat on-site disorder and on-site repulsion, and
Sec.~\ref{sm:sec:atomic-loop} the proposed atomic loop. Except in the last two
sections, we consider finite, spinless, noninteracting systems in the
grand-canonical ensemble. We set $k_B=1$ and measure energies in units of
$t_1$ for the lattice and of $E_0$ for the random-matrix model; brackets
$\langle\cdots\rangle$ denote the disorder average.

\section{Orbital response}
\label{sm:sec:response}
\label{sm:sec:reproducibility}
For a Hamiltonian $H(q)$ at temperature $T$ and chemical potential $\mu$, the
grand potential is $\Omega=\Tr g(H)$, with $g(E)=-T\ln[1+e^{-(E-\mu)/T}]$ and
$f(E)=g'(E)=[1+e^{(E-\mu)/T}]^{-1}$. Denote by $E_i$ the eigenvalues of $H$, and
by $A$ and $B$ the matrices of $H'=\partial_qH$ and $H''=\partial_q^2H$ in its
eigenbasis. Differentiating $\Omega$ twice gives the exact trace formula
\begin{align}
 -\Omega''&=-\sum_i f(E_i)B_{ii}
 -\sum_{ij}f^{[1]}(E_i,E_j)|A_{ij}|^2,
 \label{sm:eq:s_trace}\\
 f^{[1]}(x,y)&=\frac{f(x)-f(y)}{x-y},\qquad
 f^{[1]}(x,x)=f'(x).
 \label{sm:eq:s_divided}
\end{align}
The first term is the direct curvature of the Hamiltonian. The second
describes the mixing of levels; it is nonnegative, because $f$ decreases. At
charge neutrality and $q=0$ the two partners $\pm s_i$ of a chiral pair enter
the second sum twice, and together they give Eq.~(\ref{eq:pair}) of the main
text. All other terms form a remainder,
\begin{equation}
 \chi_q=\sum_i\frac{|J_i|^2}{s_i}\tanh\frac{s_i}{2T}+R_{\rm mix}+R_{\rm dir},
 \qquad R_{\rm mix}\ge0,\qquad |R_{\rm dir}|\le\|H''\|_1,
 \label{sm:eq:global-decomposition}
\end{equation}
where $R_{\rm mix}$ contains the mixing between different pairs, $R_{\rm dir}$
is the direct curvature, and the trace norm $\|\cdot\|_1$ sums the singular
values. We call the first sum the pair term, $\chi_{\rm pair}$.

We evaluate Eq.~\eqref{sm:eq:s_trace} for every flake or matrix, with the
divided difference written as
\begin{equation}
 f^{[1]}(E_i,E_j)=-f(l)[1-f(h)]\,\frac{1-e^{-(h-l)/T}}{h-l},\qquad
 h=\max(E_i,E_j),\quad l=\min(E_i,E_j),
 \label{sm:eq:stable-divided}
\end{equation}
which needs no cutoff on small gaps. As a check, independent diagonalizations,
finite differences of $\Omega$, and the comparison of symmetric and Landau gauges
agree with Eq.~\eqref{sm:eq:s_trace}. Error bars are bootstrap intervals or standard
errors over flakes and describe only the configurations present in a sample.
Because a finite sample can miss the rare flakes that dominate the mean,
Sec.~\ref{sm:sec:numerical-coefficient} integrates over them directly. The
ancillary files contain the exact crossing configuration of
Sec.~\ref{sm:sec:exact-crossing} and a script that verifies it in exact
arithmetic.

\section{Honeycomb flakes}
\label{sm:sec:lattice}
\subsection{Geometry and field coupling}
We build each flake from the hexagons with centers $(3i/2,\sqrt3(i/2+j))L_0$,
for integers $i,j$ with $\max(|i|,|j|,|i+j|)<r$. The flake has $N=6r^2$ sites,
equal sublattice populations, and open zigzag edges; $r=4$, 6, and 8 give
$N=96$, 216, and 384. Third-neighbor bonds connect opposite sublattices at
distance $2L_0$. A perpendicular field enters through
$W_{ij}(b)=W_{ij}(0)e^{i\theta_{ij}b}$, with
$\theta_{ij}=\pi(x_{A_i}y_{B_j}-y_{A_i}x_{B_j})$ in the symmetric gauge, site
coordinates in units of $L_0$, and $b=BL_0^2/\Phi_0$, $\Phi_0=2\pi\hbar/|e|$.
With the entrywise products $K=\theta\circ W$ and $K_2=-\theta^2\circ W$ at zero
field,
\begin{equation}
 H'=\begin{pmatrix}0&iK\\-iK^T&0\end{pmatrix},\qquad
 H''=\begin{pmatrix}0&K_2\\K_2^T&0\end{pmatrix}.
 \label{sm:eq:peierls-derivatives}
\end{equation}
For a simple singular value $s_i$ with real singular vectors $u_i,v_i$, the
pair amplitude is $J_i=-iu_i^TKv_i$. The physical susceptibility per flake is
$t_1(L_0^2/\Phi_0)^2\chi_b$. A gauge transformation changes both derivatives
consistently and leaves the full response unchanged.\cite{goldstein2004}

\subsection{Nearest-neighbor obstruction}
The determinant of $W$ is a sum over perfect matchings, so a zero requires
cancellation between them. Two matchings differ by alternating cycles, and
flipping a cycle of $2m$ edges changes the sign of the permutation by
$(-1)^{m-1}$.\cite{kenyon2009} In a compact flake with nearest-neighbor bonds,
let a cycle enclose $f$ hexagons, $e$ edges including its own, and $v$
interior vertices. Counting faces and using Euler's formula gives
\begin{equation}
 6f=2e-2m,\qquad v+2m-e+f=1,\qquad\text{so}\qquad v=2f-m+1.
\end{equation}
The interior vertices of an alternating cycle are matched among themselves,
so $v$ is even and $m$ odd: every flip preserves the sign. For positive
hopping, therefore,
\begin{equation}
 |\det W|=\sum_M\prod_{e\in M}W_e>0,
\end{equation}
with $M$ running over the perfect matchings and $W_e$ the hopping on bond $e$.
When the hopping amplitudes lie in a closed interval of positive values, the
smallest singular value has a positive minimum, and the mean response
saturates as $T\to0$. The argument requires every hexagon inside the
boundary to be present and gives no gap that is independent of size.

\subsection{Ensembles}
\begin{figure}[tbp]
\centering
\includegraphics[trim=162pt 0 0 0,clip,width=0.72\textwidth]{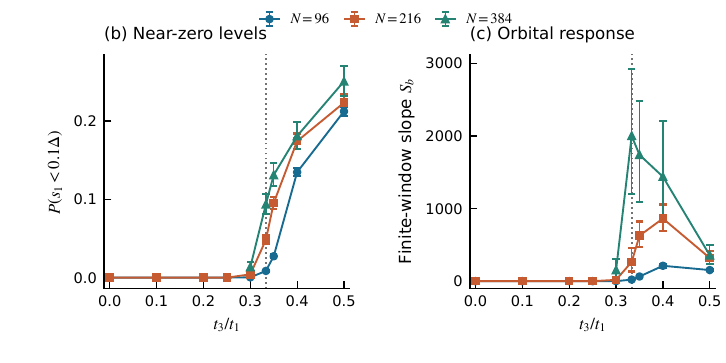}
\caption{Near-zero levels and the logarithmic slope versus third-neighbor
hopping $t_3$, for flakes of $N=96$, 216, and 384 sites with bond factors
uniform in $[0.7,1.3]$. (b) Probability that the smallest positive level $s_1$
lies below one tenth of the level spacing $\Delta$, Eq.~\eqref{sm:eq:lattice-spacing},
with 95\% Clopper--Pearson intervals. (c) Mean slope $S_b$ of the full
response versus $\ln(\Delta/T)$ over $0.003\le T/\Delta\le0.021$, in units of
$t_1$, with 95\% bootstrap intervals. Dotted lines: $t_3/t_1=1/3$, where the clean
infinite lattice acquires additional Dirac points.}
\label{sm:fig:s-lattice-comparison}
\end{figure}

Independent bond factors uniform in $[0.7,1.3]$ multiply $t_1$ and $t_3$. We use
20,007, 6,003, and 2,007 flakes for $N=96$, 216, and 384, at
$t_3/t_1=0$, 0.1, 0.2, 0.25, 0.3, 1/3, 0.35, 0.4, and 0.5 ($N=384$ from 0.3 on);
at each size the same bond factors serve all ratios. For each size and ratio
we define the level spacing
\begin{equation}
 \Delta=\frac{1}{7M}\sum_{g=1}^{M}\sum_{j=3}^{9}(s_{g,j+1}-s_{g,j}),
 \label{sm:eq:lattice-spacing}
\end{equation}
with $M$ the number of flakes and $s_{g,j}$ the $j$th positive level of flake
$g$. Figure~\ref{sm:fig:s-lattice-comparison} shows the probability of a
near-zero level, $s_1<0.1\Delta$, and the mean slope $S_b$ of the full
response versus $\ln(\Delta/T)$, fitted for each flake over the eleven
temperatures of the grid with $0.003\le T/\Delta\le0.021$. At this disorder
strength near-zero levels and a positive slope appear together, above
$t_3/t_1\approx0.3$.

\begin{figure}[tbp]
\centering\includegraphics[width=0.90\textwidth]{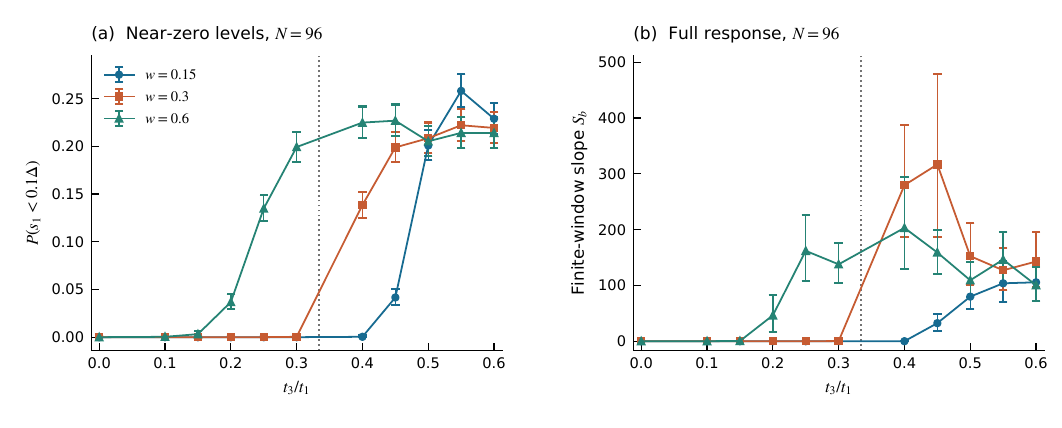}
\caption{Same as Fig.~\ref{sm:fig:s-lattice-comparison}, but for 2,500 flakes
with $N=96$ and bond factors uniform in $[1-w,1+w]$, at $w=0.15$, 0.3, and 0.6.
The same flakes serve all $w$ and $t_3/t_1$. The slope in (b) is fitted over
$0.003\le T/\Delta\le0.023$.}
\label{sm:fig:s-disorder-width}
\end{figure}

In a separate ensemble of 2,500 flakes with $N=96$ and bond factors uniform
in $[1-w,1+w]$, near-zero levels appear at smaller $t_3$ as $w$ grows
(Fig.~\ref{sm:fig:s-disorder-width}); at $w=0.6$ the number of flakes with
$s_1<0.1\Delta$ is 1, 8, and 92 at $t_3/t_1=0.1$, 0.15, and 0.2. The onset
therefore depends on disorder and is not tied to $t_3/t_1=1/3$.

\begin{figure}[tbp]
\centering\includegraphics[width=0.90\textwidth]{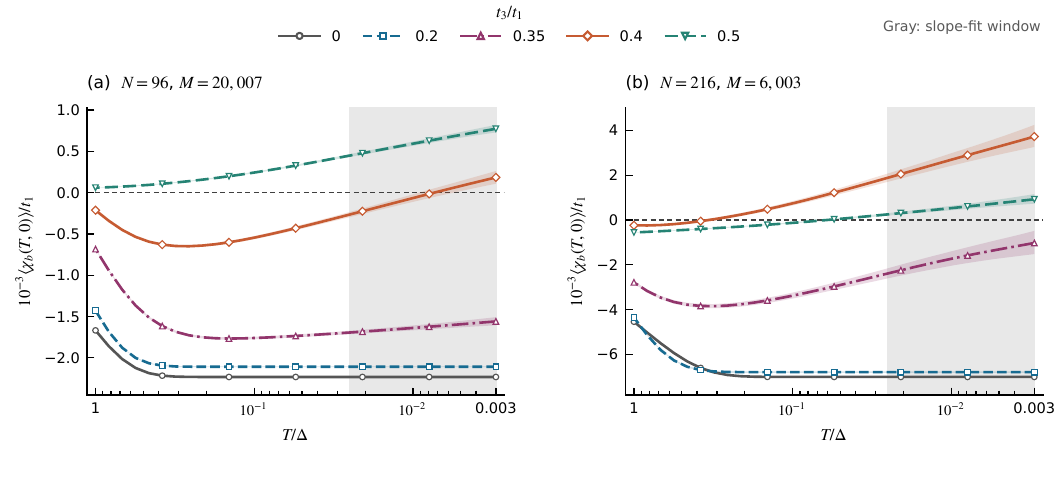}
\caption{Mean response $\langle\chi_b(T,0)\rangle$ versus
$\ln(\Delta/T)$, in units of $10^3t_1$, for (a) 20,007 flakes with $N=96$ and
(b) 6,003 flakes with $N=216$, at the values of $t_3/t_1$ indicated and
$\mu=0$. The same flakes serve all $t_3/t_1$ at a given size. Gray: the fitting
window of Fig.~\ref{sm:fig:s-lattice-comparison}(c). Bands: 95\% bootstrap
intervals.}
\label{sm:fig:s-full-temperature}
\end{figure}

Figure~\ref{sm:fig:s-full-temperature} shows the mean response itself. At
$t_3/t_1=0.35$ it stays negative over the whole temperature range for both
sizes, although its slope is already positive; at 0.4 and 0.5 it is positive
at $T/\Delta=0.003$. The mean is carried by rare flakes: removing
the 201 largest responses at $N=96$, $t_3/t_1=0.4$, and $T/\Delta=0.003$
changes it from 184.1 to $-226.6$. At $t_3/t_1=0.2$ the sample contains no
near-zero level and shows a plateau; Sec.~\ref{sm:sec:numerical-coefficient}
treats such rare events directly.

\subsection{Exact crossing}
\label{sm:sec:exact-crossing}
At $t_3/t_1=1/5$ we construct a crossing that couples to the field, in the
96-site flake of Fig.~\ref{fig:crossing} of the main text. All bond factors are
rational with the common denominator $D_0=500$, so $W=A/D_0$ with an integer
matrix $A$. We vary one nearest-neighbor entry $W_{r_0c_0}$, the purple bond.
With $d_A=\det A$ and the cofactor $C_{r_0c_0}\ne0$, the choice
\begin{equation}
 W^*_{r_0c_0}=\frac{1}{D_0}\left(A_{r_0c_0}-\frac{d_A}{C_{r_0c_0}}\right)
 \label{sm:eq:s_root}
\end{equation}
makes the determinant vanish while the cofactor stays nonzero, so $W$ has rank
47, and the cofactors $v_j=C_{r_0j}$ and $u_i=C_{ic_0}$ give the null vectors. In
exact rational arithmetic we check all 96 null equations, the bond geometry,
that every factor lies strictly inside $(0.7,1.3)$, and that the magnetic
contraction
\begin{equation}
 \sum_{ij}u_i(k_{x,A_i}k_{y,B_j}-k_{y,A_i}k_{x,B_j})W^*_{ij}v_j
\end{equation}
takes the same nonzero value in the symmetric and Landau gauges; here the
integers $(k_x,k_y)$ locate a site at $(k_x/2,\sqrt3k_y/2)L_0$. For normalized
null vectors, $h_*/t_1=1.0836915$, $g=u^T(\partial_hW)v=0.040447$, and
$j_b=u^TKv/t_1=-0.311047$. The local energies of Eq.~(\ref{eq:local}) of the
main text agree with the full spectrum to within $1.7\times10^{-6}t_1$ for
$|h-h_*|\le0.002t_1$ at $b=0$ and $10^{-4}$, while the next level stays above
$0.40t_1$.

Because the crossing is transverse, the implicit-function theorem continues
it as the other bonds vary. In a small neighborhood the next singular value
stays positive, $|J|$ stays bounded away from zero, and the disorder density
has a positive lower bound. Integrating $|J|^2\tanh(s/2T)/s$ across the crossing
surface gives $c\ln(T_*/T)$ with $c>0$. The remaining mixing terms in
Eq.~\eqref{sm:eq:s_trace} are nonnegative, and the direct term is bounded by the
trace norm of $H''$. We thus arrive at the lower bound
\begin{equation}
 \langle\chi_b(T,0)\rangle\ge c\ln(T_*/T)-c_0,
 \label{sm:eq:local-lower-bound}
\end{equation}
with constants $c>0$, $c_0$, and $T_*>0$ that do not depend on $T$. The bound
does not fix the coefficient of the logarithm; we determine it in
Sec.~\ref{sm:sec:global-theory}.

\section{Coefficient of the logarithm}
\label{sm:sec:global-theory}
\subsection{Crossing surfaces}
We express $\mathcal A(0)$ as an integral over the surface $\Sigma$, in the space
of bond factors, on which the smallest pair energy vanishes. Write
$W_e=t_ex_e$, with $t_e>0$ the hopping amplitude of bond $e$ and $x_e$ its
factor, uniform on $[0.7,1.3]$ with density $p_e=1/0.6$. Near a regular zero,
where $W$ has rank $n-1$, the signed pair energy $\varepsilon$, with
$s=|\varepsilon|$, has the gradient
\begin{equation}
 g_e=\frac{\partial\varepsilon}{\partial x_e}=t_eu_iv_j,\qquad
 G^2=\sum_eg_e^2,\qquad J=-i\sum_e\theta_ex_eg_e,
 \label{sm:eq:global-gradient}
\end{equation}
where bond $e$ connects sites $i$ and $j$ and $u,v$ are the normalized null
vectors. The gradient converts an energy interval into a volume of disorder
space, and the coarea formula gives
\begin{equation}
 \mathcal A(0)=2\int_\Sigma\frac{p(x)|J|^2}{G}\,dS
 =\sum_e\Big\langle\sum_{x_e^*}\frac{2p_e(x_e^*)|J|^2|g_e|}{G^2}\Big\rangle_{x_{-e}},
 \label{sm:eq:global-conditional}
\end{equation}
with $x_{-e}$ the factors other than $x_e$ and $x_e^*$ the zeros along $x_e$.
The factor 2 counts both sides of the surface, and the weights $g_e^2/G^2$
share each surface element among the bonds. Drawing a bond $e$ with
probability $q_e$ and averaging
\begin{equation}
 Z_e=\frac{2p_e(x_e^*)|J|^2|g_e|}{q_eG^2}
 \label{sm:eq:conditional-root-estimator}
\end{equation}
over the other factors gives an unbiased estimate of $\mathcal A(0)$. Since
$\det W$ is affine in $x_e$, one linear solve locates the zero: for $Wy=e_i$,
with $e_i$ the $i$th unit vector, $x_e^*=x_e-1/(t_ey_j)$.

\subsection{Finiteness}
Crossings with a small gradient $G$ have a large weight $1/G$ in
Eq.~\eqref{sm:eq:global-conditional}, but their magnetic amplitude is small
as well: Eq.~\eqref{sm:eq:global-gradient} gives $|J|^2\le L^2G^2$, with $L^2=\max_x\sum_e\theta_e^2x_e^2$ on the box of bond
factors; the same bound holds for any simple positive singular value with
$g_e=\partial s/\partial x_e$. Since $|g_e|\le t_e$, each term in
Eq.~\eqref{sm:eq:global-conditional} obeys
\begin{equation}
 p_e\frac{|J|^2|g_e|}{G^2}\le p_eL^2t_e.
 \label{sm:eq:global-domination}
\end{equation}
This bound relies on the Peierls form of the field coupling. At fixed $s$,
$\det(H-s)$ is at most quadratic in each $x_e$, so a line along one factor
crosses the level $s$ at most twice, up to a set of measure zero. For almost
every choice of the other factors, the line either contains a regular zero of
$\det W$ or keeps a gap. Dominated convergence then gives
\begin{equation}
 \lim_{s\to0}\mathcal A(s)=\mathcal A(0)\le2L^2\sum_e\|p_e\|_\infty t_e<\infty.
 \label{sm:eq:global-pair-theorem}
\end{equation}
At a zero, a gauge transformation changes $K$ by $\Lambda_AW-W\Lambda_B$, with
diagonal $\Lambda_A,\Lambda_B$, and this change drops out of the contraction
with the null vectors, so $\mathcal A(0)$ is gauge invariant. The exact crossing
of Sec.~\ref{sm:sec:exact-crossing} gives $\mathcal A(0)>0$ at $N=96$ and
$t_3/t_1=1/5$; for positive nearest-neighbor hopping on a compact flake,
$\mathcal A(0)=0$.

The finite limit fixes the slope of the mean pair term,
$d\langle\chi_{\rm pair}\rangle/d\ln(1/T)\to\mathcal A(0)$. The same coefficient
governs the full response if the remainder in
Eq.~\eqref{sm:eq:global-decomposition} grows more slowly than $\ln(1/T)$. A
sufficient condition is $\langle1/s_{(2)}\rangle<\infty$ for the second-smallest
level $s_{(2)}$, since $0\le R_{\rm mix}\le\|H'\|_F^2/s_{(2)}$, with $\|\cdot\|_F$ the
Frobenius norm. We have not proved this condition;
Sec.~\ref{sm:sec:numerical-coefficient} tests the full response directly.

\subsection{Numerical evaluation}
\label{sm:sec:crossing-coefficient}
At $N=96$ and $t_3/t_1=0.4$, 524,288 lines along uniformly chosen bonds contain
59,880 zeros and give $\mathcal A(0)=231.5$, with 95\% bootstrap interval
$[227.3,235.7]$. Sharing the surface among bonds with weights proportional
to $|g_e|$ instead gives 232.3 on the same zeros, and an independent
sample with $q_e\propto t_e^2$ gives 233.1. Table~\ref{sm:tab:crossing-size} lists
the three sizes: $\mathcal A(0)$ grows with $N$, while $\mathcal A(0)$ per unit
area does not change monotonically, and three sizes do not determine a
scaling law. At $t_3/t_1=0.2$, 131,072 lines contain no zero: the exact crossing
proves $\mathcal A(0)>0$ there, but its magnitude lies below the resolution of
this sample.

\begin{table}[tbp]
\centering
\caption{Coefficient $\mathcal A(0)$ at $t_3/t_1=0.4$, in units of $t_1$, with 95\%
bootstrap intervals, for bond factors uniform in $[0.7,1.3]$;
$a_{\rm hull}$ is the area of the convex hull of the sites in units of $L_0^2$.}
\label{sm:tab:crossing-size}
\begin{tabular}{rrrrr}
\toprule
$N$ & $a_{\rm hull}$ & $\mathcal A(0)$ & 95\% interval & $\mathcal A(0)/a_{\rm hull}$\\
\midrule
96 & 103.9 & 231.5 & [227.3, 235.7] & 2.23\\
216 & 249.4 & 803.8 & [769.5, 836.9] & 3.22\\
384 & 457.3 & 1352.7 & [1301.5, 1407.9] & 2.96\\
\bottomrule
\end{tabular}
\end{table}

\subsection{Complete response}
\label{sm:sec:numerical-coefficient}
A finite sample saturates below its smallest gap, $3.3\times10^{-7}t_1$ for the
20,007 flakes at $N=96$ and $t_3/t_1=0.4$. To resolve smaller gaps we integrate
continuously along one bond factor. With $g_e=\partial s_1/\partial x_e$ for the
lowest pair and $\omega_e=g_e^2/\sum_fg_f^2$, which sum to one, any observable $O$
obeys
\begin{equation}
 \langle O\rangle=\Big\langle\int_{0.7}^{1.3}\frac{p_e(x_e)\omega_e(x)}{q_e}\,O(x)\,dx_e\Big\rangle_{e,\,x_{-e}}.
 \label{sm:eq:thermal-conditional}
\end{equation}
Because the weights sum to one, this holds for the full response, including
all levels and the direct curvature. We integrate along 2,048 lines
on uniformly chosen bonds, 234 of which contain a zero, with quadrature
panels that resolve the thermal width $T/|g_e|$ around each zero, down to
$T/\Delta=10^{-8}$.

\begin{figure}[tbp]
\centering\includegraphics[width=\textwidth]{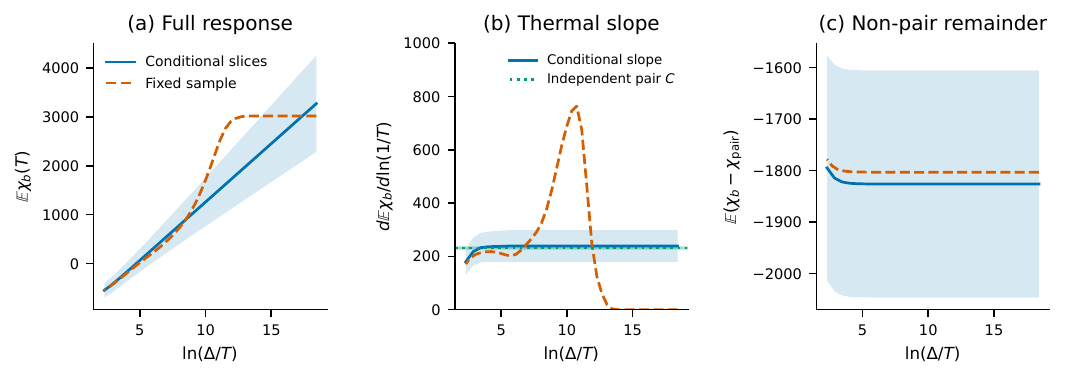}
\caption{Full response for $N=96$, $t_3/t_1=0.4$, and $\mu=0$, in units of
$t_1$. Blue: integration along 2,048 lines, Eq.~\eqref{sm:eq:thermal-conditional};
orange dashes: the fixed sample of 20,007 flakes. (a) Mean response, (b) its
slope versus $\ln(\Delta/T)$, compared with $\mathcal A(0)=231.5$ (green), and
(c) the remainder $\chi_b-\chi_{\rm pair}$. Bands: 1.96 standard errors;
$\Delta=0.031946t_1$.}
\label{sm:fig:thermal}
\end{figure}

The slope of the mean response approaches $238.7\pm30.6$ (standard error), in
agreement with $\mathcal A(0)=231.5\pm2.2$, while the mean remainder
$\chi_b-\chi_{\rm pair}$ approaches $-1826\pm113$ and changes by less than
$4\times10^{-11}$ over the lowest decade of temperature
(Fig.~\ref{sm:fig:thermal}). Within the statistical error, the logarithm of
the full response therefore has the coefficient $\mathcal A(0)$.

\section{Random-matrix model}
\label{sm:sec:gaussian}
\subsection{Density identity}
We consider $W(a)=(X+iaY)/\sqrt{1+a^2}$, with $X$ and $Y$ independent real
$(n+\nu)\times n$ matrices of Gaussian entries with zero mean and variance $1/n$.
At $a=0$, $W'=iY$ and $W''=-X$. Write $X=USV^T$, with $S$ containing the positive
singular values $s_i$ and $U,V$ orthogonal. At fixed $X$ the entries of
$Z=U^TYV$ are again independent Gaussians of variance $1/n$. Second-order
perturbation theory gives the level curvature
\begin{equation}
 K_i=s_i''(0)=-s_i+\frac{Z_{ii}^2+\sum_{\alpha=n+1}^{n+\nu}Z_{\alpha i}^2}{s_i}
 +\frac12\sum_{j\ne i}\left[\frac{(Z_{ij}-Z_{ji})^2}{s_i-s_j}
 +\frac{(Z_{ij}+Z_{ji})^2}{s_i+s_j}\right],
 \label{sm:eq:s_K}
\end{equation}
and its average over $Y$,
\begin{equation}
 \overline K_i=-s_i+\frac{\nu+1}{ns_i}
 +\frac1n\sum_{j\ne i}\left(\frac1{s_i-s_j}+\frac1{s_i+s_j}\right).
 \label{sm:eq:s_conditional}
\end{equation}
The three groups of terms are the direct curvature, the mixing with the
partner and with the $\nu$ zero modes, and the mixing with the other pairs. The
first-order level shifts vanish because $H(0)$ is real, so
$\chi_a=\sum_iq_\mu(s_i)K_i$ with $q_\mu(s)=f(-s)-f(s)$. The positive levels have
the joint density
\begin{equation}
 p(s_1,\dots,s_n)\propto e^{-(n/2)\sum_is_i^2}\prod_is_i^\nu\prod_{i<j}|s_i^2-s_j^2|,
 \label{sm:eq:s_jpdf}
\end{equation}
and one readily verifies that $n\overline K_i=\partial_{s_i}\ln p+1/s_i$.
Multiplying by $q_\mu(s_i)$ and integrating by parts, without boundary terms
because $q_\mu(0)=0$ and $p$ vanishes at collisions and at infinity, we arrive
at Eq.~(\ref{eq:identity}) of the main text,
\begin{equation}
 \langle\chi_a(T,0;\mu)\rangle=\frac1n\int_0^\infty ds\,R_{n,\nu}(s)
 \left[\frac{q_\mu(s)}s-\frac{dq_\mu}{ds}\right],
 \label{sm:eq:s_identity}
\end{equation}
with $R_{n,\nu}$ the density of positive levels, normalized to $n$. The identity
holds after the disorder average; for a single matrix its right-hand side is
an unbiased estimate of the mean.

\subsection{Coefficient}
The normalization of Eq.~\eqref{sm:eq:s_jpdf} follows from the Laguerre--Selberg
integral,\cite{selberg2008,wirtz2015}
\begin{equation}
 Z_{n,\nu}=2^{-n}\Big(\frac n2\Big)^{-n(n+\nu)/2}
 \prod_{j=1}^n\frac{\Gamma(1+j/2)\,\Gamma[(\nu+j)/2]}{\Gamma(3/2)}.
 \label{sm:eq:s_Z}
\end{equation}
When one level approaches zero, each factor $|s_j^2-s_1^2|$ tends to $s_j^2$, so
the other $n-1$ levels have the density of index $\nu+2$ at the same
confinement. Hence
\begin{equation}
 R_{n,\nu}(s)=A_{n,\nu}s^\nu[1+O(s^2)],\qquad
 A_{n,\nu}=n\frac{Z_{n-1,\nu+2}}{Z_{n,\nu}}.
 \label{sm:eq:s_A}
\end{equation}
For $\nu=0$ the constant density produces the logarithm,
\begin{equation}
 \langle\chi_a(T,0)\rangle=C_n\ln(1/T)+K_n+o(1),\qquad
 C_n=\frac{A_{n,0}}n=\sqrt{\frac2n}\,\frac{\Gamma[(n+1)/2]}{\Gamma(n/2)}
 =1-\frac1{4n}+O(n^{-2}),
 \label{sm:eq:s_gaussian_log}
\end{equation}
with a temperature-independent constant $K_n$; the term $-dq_\mu/ds$ contributes
$-C_n$ to $K_n$ and nothing to the logarithm. For $\nu>0$ the integrand behaves
as $s^{\nu-1}$ near zero, and the response saturates.

\begin{figure}[tbp]
\centering\includegraphics[width=0.90\textwidth]{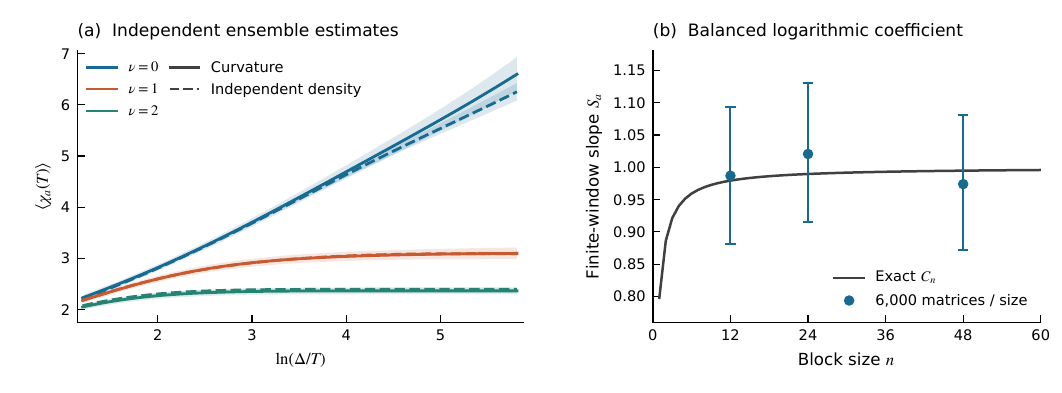}
\caption{Check of the random-matrix result at $\mu=0$, with $\Delta=\pi/(2n)$.
(a) Mean response from the curvatures (solid) and from the density identity
(dashed), each from 6,000 matrices with $n=24$ and $\nu=0,1,2$. (b) Slopes $S_a$
of the mean response versus $\ln(\Delta/T)$, fitted over
$0.005\le T/\Delta\le0.025$, and the coefficient $C_n$. Bands and bars: 95\%
bootstrap intervals.}
\label{sm:fig:s-gaussian-validation}
\end{figure}

Figure~\ref{sm:fig:s-gaussian-validation}(a) compares two estimates of the
mean for $n=24$, from the curvatures~\eqref{sm:eq:s_K} and from the
identity~\eqref{sm:eq:s_identity}, each on its own 6,000 matrices, and
Fig.~\ref{sm:fig:s-gaussian-validation}(b) compares fitted slopes for $n=12$, 24,
and 48 with $C_n$. Here $\Delta=\pi/(2n)$ is the spacing from the macroscopic
level density; it differs from $1/R_{n,0}(0)=1/(nC_n)$.

\subsection{Chemical-potential cutoff}
\label{sm:sec:s_detuning}
Write $m=\mu/T$ and $x=s/T$, so that $q_m(x)=\sinh x/(\cosh x+\cosh m)$.
Subtracting Eq.~\eqref{sm:eq:s_identity} at $\mu=0$ cancels the derivative
terms, because $\int_0^\infty(dq_m/dx)dx=1$ for every $m$. For $T$ and $|\mu|$
below the energy at which $R_{n,0}$ departs from its zero-energy value, this
leaves
\begin{equation}
 \langle\chi_a(T,0;\mu)\rangle-\langle\chi_a(T,0;0)\rangle=C_nG(\mu/T),\qquad
 G(m)=\int_0^\infty\frac{q_m(x)-q_0(x)}x\,dx
 =\psi(\tfrac12)-\operatorname{Re}\psi\!\left(\tfrac12+\tfrac{im}{2\pi}\right),
 \label{sm:eq:s_detuning_G}
\end{equation}
with $\psi$ the digamma function. The function $G$ is even and nonpositive,
$G(m)=O(m^2)$ for $|m|\ll1$, and $G(m)=-\ln|m|+\ln(\pi/2)-\gamma+o(1)$ for
$|m|\gg1$, with Euler's constant $\gamma$. The logarithm thus stops growing once
$T\ll|\mu|$, at $C_n\ln(1/|\mu|)$ plus a constant.

\subsection{Sign reversal along the crossover}
At $\mu=0$ and fixed $T>0$, let $F(a)=\langle\Omega(a)\rangle$ and
$\overline\chi_a(a)=-F''(a)$, the mean differential response at finite $a$.
Derivative and average commute, because $H'$ and $H''$ grow at most linearly
in the Gaussian entries. Complex conjugation makes $F$ even in $a$. For $a>0$,
$W(1/a;Y,X)=iW(a;X,Y)^*$ has the same singular values as $W(a;X,Y)$, and $X$
and $Y$ are identically distributed, so
\begin{equation}
 F(a)=F(1/a),\qquad F'(0)=F'(1)=0,\qquad\int_0^1\overline\chi_a(a)\,da=0.
 \label{sm:eq:finite-gaussian-integral}
\end{equation}
At $a=0$ the kernel of Eq.~\eqref{sm:eq:s_identity} is $h(s/2T)/s$, with
$h(x)=\tanh x-x\operatorname{sech}^2x>0$ for $x>0$, so $\overline\chi_a(0)>0$ for
every $n$, $\nu$, and $T$. The mean response must therefore be negative on an
interval inside $0<a<1$. The Peierls field has no counterpart of the symmetry
$a\to1/a$; for the lattice, Sec.~\ref{sm:sec:finite-field} gives the calculation.

\section{Finite field}
\label{sm:sec:finite-field}
Figure~\ref{fig:field} of the main text uses the 20,007 flakes with $N=96$ and
$t_3/t_1=0.4$ of Sec.~\ref{sm:sec:lattice}, at $T=0.003\Delta$ and
$b=0$, $10^{-5}$, $3\times10^{-5}$, $10^{-4}$, and $3\times10^{-4}$. The zero-field
mean, 184.1, coincides with that of Fig.~\ref{sm:fig:s-full-temperature}, and
the mean differential response changes sign between
\begin{equation}
 \overline\chi_b(3\times10^{-5})=103.4\ [44.1,164.6],\qquad
 \overline\chi_b(10^{-4})=-143.2\ [-177.0,-108.3],
\end{equation}
with 95\% bootstrap intervals; these fields correspond to fluxes of
$0.0031\Phi_0$ and $0.0104\Phi_0$ through the flake. The differential response
is the field derivative of the moment $M_b=-\partial_b\Omega$, so a negative value
means that the moment decreases with field, whatever its sign.

For an isolated pair, let $\delta$ be its signed zero-field energy and $J$ its
magnetic amplitude, so that its levels are
$\pm\varepsilon$ with $\varepsilon=(\delta^2+J^2b^2)^{1/2}$. Differentiating
$\Omega_{\rm pair}=-2T\ln[2\cosh(\varepsilon/2T)]$ gives
\begin{equation}
 \chi_{\rm pair}=\frac{J^2\delta^2}{\varepsilon^3}\tanh\frac{\varepsilon}{2T}
 +\frac{J^4b^2}{2T\varepsilon^2}\operatorname{sech}^2\frac{\varepsilon}{2T}\ge0.
 \label{sm:eq:audit-pair}
\end{equation}
At $\delta=0$ this is $J^2\operatorname{sech}^2(|J|b/2T)/2T$, so the field suppresses
the pair response without reversing it, and the negative contribution must
come from the rest of the spectrum. For the exact crossing of
Sec.~\ref{sm:sec:exact-crossing} at $T=10^{-5}t_1$, the full zero-field
response is 3741.3, of which the pair gives 4837.5; at $b=2.35\times10^{-4}$ the
full response is $-1003.9$ while the moment is still positive.

\section{On-site disorder}
\label{sm:sec:robustness}
\begin{figure}[tbp]
\centering
\includegraphics[width=\textwidth]{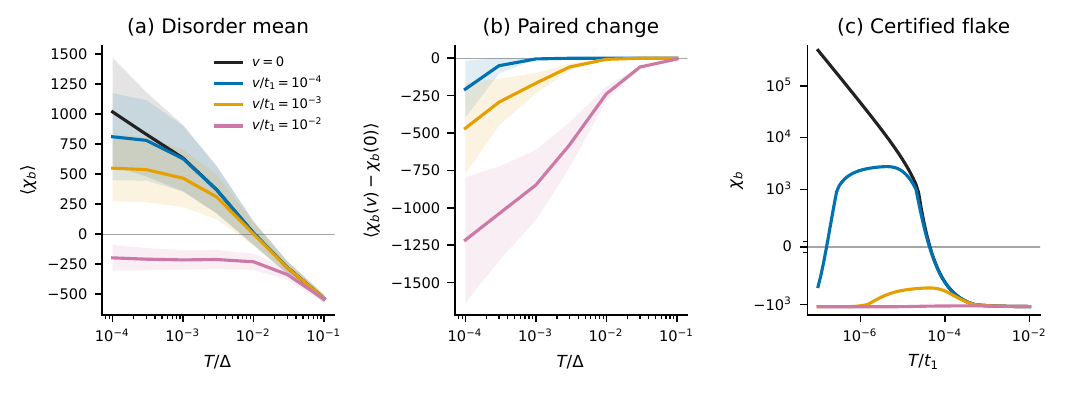}
\caption{Effect of on-site energies of half-width $v$ at $\mu=0$, in units of
$t_1$. (a) Mean response of the 8,192 flakes of Table~\ref{sm:tab:onsite-control}
and (b) its change relative to $v=0$ on the same flakes; bands: 1.96 standard
errors, $\Delta=0.031946t_1$. (c) One on-site pattern applied to the exact
crossing at $t_3/t_1=0.2$, versus $T/t_1$; the vertical scale is linear for
$|\chi_b|<10^3$ and logarithmic beyond.}
\label{sm:fig:onsite-robustness}
\end{figure}

On-site energies break the chiral pairing and can move a small pair away
from the chemical potential. We add $V=v\sum_iz_i|i\rangle\langle i|$, with
independent $z_i$ uniform in $[-1,1]$ and half-width $v$, keep $\mu=0$, and
evaluate the response, Eq.~\eqref{sm:eq:s_trace}, of 8,192 new flakes with
$N=96$ and $t_3/t_1=0.4$; the same hopping factors and site energies serve all
$v$ and $T$. At $T=0.003\Delta$ the mean stays positive for $v=10^{-3}t_1$ and
turns negative for $v=10^{-2}t_1$ (Table~\ref{sm:tab:onsite-control} and
Fig.~\ref{sm:fig:onsite-robustness}). The median is negative even at $v=0$: two
thirds of the flakes are diamagnetic, and the positive mean comes from the
minority with a nearly closed pair. Applied to the exact crossing, an on-site
pattern with $v=10^{-3}t_1$ moves both central levels above $\mu$, to
$2.9\times10^{-6}t_1$ and $2.4\times10^{-4}t_1$, which removes the pair
contribution. Same-sublattice hopping, correlated potentials, and a gate that
restores half filling require separate calculations.

\begin{table}[tbp]
\centering
\caption{Response at $T=0.003\Delta$ and $\mu=0$, in units of $t_1$, for
8,192 flakes with $N=96$ and $t_3/t_1=0.4$ and on-site energies of half-width
$v$: mean with standard error, median, and fraction of flakes with
$\chi_b>0$.}
\label{sm:tab:onsite-control}
\begin{tabular}{rrrr}
\toprule
$v/t_1$ & Mean & Median & $\chi_b>0$\\
\midrule
$0$ & $369\pm100$ & $-624$ & 0.32\\
$10^{-4}$ & $368\pm100$ & $-624$ & 0.32\\
$10^{-3}$ & $310\pm94$ & $-624$ & 0.32\\
$10^{-2}$ & $-211\pm40$ & $-644$ & 0.31\\
\bottomrule
\end{tabular}
\end{table}

\section{On-site interaction}
\label{sm:sec:interaction}
On-site repulsion matters most for a nearly closed pair, whose small
splitting $s$ cannot compete with the interaction energy of its two orbitals.
We estimate the resulting cutoff by projecting a Hubbard interaction of
strength $U$ onto these orbitals, $\alpha=(u,0)$ on sublattice A and
$\beta=(0,v)$ on sublattice B, with $Wv=su$ at zero field. With spin, the
projected Hamiltonian is
\begin{equation}
 H_{\rm pair}=\sum_{\sigma}\bigl(\kappa\,c^\dagger_{\alpha\sigma}c_{\beta\sigma}+{\rm h.c.}\bigr)
 +\sum_{x=\alpha,\beta}U_x\bigl(n_{x\uparrow}-\tfrac12\bigr)\bigl(n_{x\downarrow}-\tfrac12\bigr),
 \label{sm:eq:hubbard-pair}
\end{equation}
with $\kappa=s+iJb$, $U_\alpha=U\sum_iu_i^4$, and $U_\beta=U\sum_jv_j^4$. The two
orbitals lie on different sublattices, so the on-site interaction does not
couple them. For $U_\alpha=U_\beta=U_{\rm eff}$ and two electrons in the pair, the
ground state is a singlet with orbital response
\begin{equation}
 \chi_{\rm pair}(T=0)=\frac{2|J|^2}{\sqrt{s^2+U_{\rm eff}^2/16}}.
 \label{sm:eq:hubbard-dimer}
\end{equation}
At $U_{\rm eff}=0$ this is twice Eq.~(\ref{eq:pair}) at $T=0$, one term per spin.
The divergence at $s\to0$ is cut off at $s\simeq U_{\rm eff}/4$. The
singlet--triplet splitting, $(U_{\rm eff}^2/4+4s^2)^{1/2}-U_{\rm eff}/2\simeq4s^2/U_{\rm eff}$,
vanishes as the pair closes, so a nearly closed pair carries two nearly free
spins.

For the lowest pair of every flake in the $t_3/t_1=0.4$ ensembles we take
$U_{\rm eff}=U(\sum_iu_i^4+\sum_jv_j^4)/2$ and solve
Eq.~\eqref{sm:eq:hubbard-pair} exactly in its 16-dimensional Fock space at
$\mu=0$ and finite $T$; at $U=0$ this reproduces the spinful pair term
$2|J|^2\tanh(s/2T)/s$. Among the pairs with $s_1<0.1\Delta$ the median
$U_{\rm eff}/\Delta$ is 2.1, 2.6, and 3.1 for $N=96$, 216, and 384 at $U=t_1$, so
the cutoff $U_{\rm eff}/4$ lies at $0.5$--$0.8\,\Delta$ and grows with size.
At $U=t_1/2$ the slope of the mean pair response already drops by an order of
magnitude (Table~\ref{sm:tab:hubbard}), and the spin response of the same pairs
exceeds the orbital one by a factor of 26 to 48 at $T=0.003\Delta$, for
$t_1=2.7$~eV, $L_0=0.142$~nm, and $g=2$. The estimate neglects screening, the
coupling to other levels, and the difference between $U_\alpha$ and $U_\beta$.

\begin{table}[tbp]
\centering
\caption{Slope of the mean spinful pair response versus $\ln(\Delta/T)$, fitted
over $0.003\le T/\Delta\le0.025$, for the projected Hubbard pair,
Eq.~\eqref{sm:eq:hubbard-pair}, at $t_3/t_1=0.4$, in units of $t_1$. At $U=0$ the
slope is close to $2\mathcal A(0)$.}
\label{sm:tab:hubbard}
\begin{tabular}{rrrrr}
\toprule
$N$ & $U=0$ & $U=t_1/2$ & $U=t_1$ & $U=2t_1$\\
\midrule
96 & 427 & 39 & 28 & 20\\
216 & 1729 & 141 & 98 & 66\\
384 & 2897 & 193 & 119 & 79\\
\bottomrule
\end{tabular}
\end{table}

\section{Atomic loop}
\label{sm:sec:atomic-loop}
We consider two noninteracting spin-polarized fermions on a square with
positive bond amplitudes $u_1,u_2,u_3,u_4$ in cyclic order and equal site
energies, set to zero. The one-particle Hamiltonian has the block form of
Eq.~(\ref{eq:chiral}) of the main text, with
\begin{equation}
 W(\theta)=\begin{pmatrix}
 u_1e^{i\theta/4}&u_4e^{-i\theta/4}\\
 u_2e^{-i\theta/4}&u_3e^{i\theta/4}
 \end{pmatrix},\qquad
 \det W=u_1u_3e^{i\theta/2}-u_2u_4e^{-i\theta/2},
 \label{sm:eq:atomic-loop-block}
\end{equation}
where $\theta$ is the total Peierls phase around the loop. With
$D=u_1u_3-u_2u_4$ and $S=(\sum_iu_i^2)^{1/2}$, the spectrum is $\pm s_1,\pm s_2$,
\begin{equation}
 s_{1,2}^2=\tfrac12\Bigl[S^2\mp\bigl(S^4-4|\det W|^2\bigr)^{1/2}\Bigr].
 \label{sm:eq:atomic-loop-spectrum}
\end{equation}
At $D=\theta=0$ the lower level vanishes and the upper one equals $S$. Since
$\partial D/\partial u_1=u_3>0$, this is a regular crossing surface. With
$P=u_1u_3=u_2u_4$ at the crossing, the local pair energies are
\begin{equation}
 E_\pm\simeq\pm\frac{(D^2+P^2\theta^2)^{1/2}}{S},\qquad |J_\theta|=\frac PS,
 \label{sm:eq:atomic-loop-local}
\end{equation}
the analog of Eq.~(\ref{eq:local}) of the main text. Bond amplitudes drawn
independently and uniformly from $[u_0(1-w),u_0(1+w)]$, with $0<w<1$, sample
both sides of the surface, while the lower bound $u_0(1-w)>0$ keeps the upper
pair away from zero; the mean response then grows logarithmically.

With two particles, the six many-body states give the exact partition
function
\begin{equation}
 Z_2=2+4\cosh(s_1/T)\cosh(s_2/T),\qquad F_2=-T\ln Z_2,
 \label{sm:eq:atomic-loop-partition}
\end{equation}
and the response to the synthetic flux is $\chi_\theta=-\partial_\theta^2F_2$ at
$\theta=0$; the circulating current is $I_\theta=-\partial_\theta F_2/\hbar$. For
$T\ll S$ near the crossing, one fermion occupies the level near $-S$ and the
other the central pair, so the singular part of the response is
\begin{equation}
 \chi_\theta\simeq\frac{|J_\theta|^2}{s_1}\tanh\frac{s_1}{T},
 \label{sm:eq:canonical-pair}
\end{equation}
plus a bounded correction from the remote levels. The canonical factor
$\tanh(s_1/T)$ replaces the grand-canonical $\tanh(s_1/2T)$ of
Eq.~(\ref{eq:pair}) and changes only the thermal crossover, not the coefficient
of the logarithm. As a check, the spectrum, the partition function, and the
flux curvature agree with direct diagonalization and enumeration of the
two-particle states.

Tunnel-coupled fermionic tweezer plaquettes\cite{yan2022} and chiral edge
states of fermions in a synthetic gauge field\cite{mancini2015} demonstrate
separate ingredients of this proposal. The test would combine a loop flux with
repeated draws of the bond amplitudes at fixed half filling. It requires
site-energy differences and residual interactions below the pair energies to
be resolved; these conditions and the attainable temperature must be
established for a specific implementation.


\begin{thebibliography}{30}%
\makeatletter
\providecommand \@ifxundefined [1]{%
 \@ifx{#1\undefined}
}%
\providecommand \@ifnum [1]{%
 \ifnum #1\expandafter \@firstoftwo
 \else \expandafter \@secondoftwo
 \fi
}%
\providecommand \@ifx [1]{%
 \ifx #1\expandafter \@firstoftwo
 \else \expandafter \@secondoftwo
 \fi
}%
\providecommand \natexlab [1]{#1}%
\providecommand \enquote  [1]{``#1''}%
\providecommand \bibnamefont  [1]{#1}%
\providecommand \bibfnamefont [1]{#1}%
\providecommand \citenamefont [1]{#1}%
\providecommand \href@noop [0]{\@secondoftwo}%
\providecommand \href [0]{\begingroup \@sanitize@url \@href}%
\providecommand \@href[1]{\@@startlink{#1}\@@href}%
\providecommand \@@href[1]{\endgroup#1\@@endlink}%
\providecommand \@sanitize@url [0]{\catcode `\\12\catcode `\$12\catcode
  `\&12\catcode `\#12\catcode `\^12\catcode `\_12\catcode `\%12\relax}%
\providecommand \@@startlink[1]{}%
\providecommand \@@endlink[0]{}%
\providecommand \url  [0]{\begingroup\@sanitize@url \@url }%
\providecommand \@url [1]{\endgroup\@href {#1}{\urlprefix }}%
\providecommand \urlprefix  [0]{URL }%
\providecommand \Eprint [0]{\href }%
\providecommand \doibase [0]{https://doi.org/}%
\providecommand \selectlanguage [0]{\@gobble}%
\providecommand \bibinfo  [0]{\@secondoftwo}%
\providecommand \bibfield  [0]{\@secondoftwo}%
\providecommand \translation [1]{[#1]}%
\providecommand \BibitemOpen [0]{}%
\providecommand \bibitemStop [0]{}%
\providecommand \bibitemNoStop [0]{.\EOS\space}%
\providecommand \EOS [0]{\spacefactor3000\relax}%
\providecommand \BibitemShut  [1]{\csname bibitem#1\endcsname}%
\let\auto@bib@innerbib\@empty
\bibitem [{\citenamefont {Castro~Neto}\ \emph {et~al.}(2009)\citenamefont
  {Castro~Neto}, \citenamefont {Guinea}, \citenamefont {Peres}, \citenamefont
  {Novoselov},\ and\ \citenamefont {Geim}}]{castroNeto2009}%
  \BibitemOpen
  \bibfield  {author} {\bibinfo {author} {\bibfnamefont {A.~H.}\ \bibnamefont
  {Castro~Neto}}, \bibinfo {author} {\bibfnamefont {F.}~\bibnamefont {Guinea}},
  \bibinfo {author} {\bibfnamefont {N.~M.~R.}\ \bibnamefont {Peres}}, \bibinfo
  {author} {\bibfnamefont {K.~S.}\ \bibnamefont {Novoselov}},\ and\ \bibinfo
  {author} {\bibfnamefont {A.~K.}\ \bibnamefont {Geim}},\ }\bibfield  {title}
  {\bibinfo {title} {The electronic properties of graphene},\ }\href
  {https://doi.org/10.1103/RevModPhys.81.109} {\bibfield  {journal} {\bibinfo
  {journal} {Rev. Mod. Phys.}\ }\textbf {\bibinfo {volume} {81}},\ \bibinfo
  {pages} {109} (\bibinfo {year} {2009})},\ \Eprint
  {https://arxiv.org/abs/0709.1163} {arXiv:0709.1163} \BibitemShut {NoStop}%
\bibitem [{\citenamefont {McClure}(1956)}]{mcclure1956}%
  \BibitemOpen
  \bibfield  {author} {\bibinfo {author} {\bibfnamefont {J.~W.}\ \bibnamefont
  {McClure}},\ }\bibfield  {title} {\bibinfo {title} {Diamagnetism of
  graphite},\ }\href {https://doi.org/10.1103/PhysRev.104.666} {\bibfield
  {journal} {\bibinfo  {journal} {Phys. Rev.}\ }\textbf {\bibinfo {volume}
  {104}},\ \bibinfo {pages} {666} (\bibinfo {year} {1956})}\BibitemShut
  {NoStop}%
\bibitem [{\citenamefont {Fukuyama}(2007)}]{fukuyama2007}%
  \BibitemOpen
  \bibfield  {author} {\bibinfo {author} {\bibfnamefont {H.}~\bibnamefont
  {Fukuyama}},\ }\bibfield  {title} {\bibinfo {title} {Anomalous orbital
  magnetism and {Hall} effect of massless fermions in two dimension},\ }\href
  {https://doi.org/10.1143/JPSJ.76.043711} {\bibfield  {journal} {\bibinfo
  {journal} {J. Phys. Soc. Jpn.}\ }\textbf {\bibinfo {volume} {76}},\ \bibinfo
  {pages} {043711} (\bibinfo {year} {2007})},\ \Eprint
  {https://arxiv.org/abs/cond-mat/0703010} {arXiv:cond-mat/0703010}
  \BibitemShut {NoStop}%
\bibitem [{\citenamefont {Vallejo~Bustamante}\ \emph
  {et~al.}(2021)\citenamefont {Vallejo~Bustamante}, \citenamefont {Wu},
  \citenamefont {Fermon}, \citenamefont {Pannetier-Lecoeur}, \citenamefont
  {Wakamura}, \citenamefont {Watanabe}, \citenamefont {Taniguchi},
  \citenamefont {Pellegrin}, \citenamefont {Bernard}, \citenamefont
  {Daddinounou}, \citenamefont {Bouchiat}, \citenamefont {Gu{\'e}ron},
  \citenamefont {Ferrier}, \citenamefont {Montambaux},\ and\ \citenamefont
  {Bouchiat}}]{vallejo2021}%
  \BibitemOpen
  \bibfield  {author} {\bibinfo {author} {\bibfnamefont {J.}~\bibnamefont
  {Vallejo~Bustamante}}, \bibinfo {author} {\bibfnamefont {N.~J.}\ \bibnamefont
  {Wu}}, \bibinfo {author} {\bibfnamefont {C.}~\bibnamefont {Fermon}}, \bibinfo
  {author} {\bibfnamefont {M.}~\bibnamefont {Pannetier-Lecoeur}}, \bibinfo
  {author} {\bibfnamefont {T.}~\bibnamefont {Wakamura}}, \bibinfo {author}
  {\bibfnamefont {K.}~\bibnamefont {Watanabe}}, \bibinfo {author}
  {\bibfnamefont {T.}~\bibnamefont {Taniguchi}}, \bibinfo {author}
  {\bibfnamefont {T.}~\bibnamefont {Pellegrin}}, \bibinfo {author}
  {\bibfnamefont {A.}~\bibnamefont {Bernard}}, \bibinfo {author} {\bibfnamefont
  {S.}~\bibnamefont {Daddinounou}}, \bibinfo {author} {\bibfnamefont
  {V.}~\bibnamefont {Bouchiat}}, \bibinfo {author} {\bibfnamefont
  {S.}~\bibnamefont {Gu{\'e}ron}}, \bibinfo {author} {\bibfnamefont
  {M.}~\bibnamefont {Ferrier}}, \bibinfo {author} {\bibfnamefont
  {G.}~\bibnamefont {Montambaux}},\ and\ \bibinfo {author} {\bibfnamefont
  {H.}~\bibnamefont {Bouchiat}},\ }\bibfield  {title} {\bibinfo {title}
  {Detection of graphene's divergent orbital diamagnetism at the {Dirac}
  point},\ }\href {https://doi.org/10.1126/science.abf9396} {\bibfield
  {journal} {\bibinfo  {journal} {Science}\ }\textbf {\bibinfo {volume}
  {374}},\ \bibinfo {pages} {1399} (\bibinfo {year} {2021})},\ \Eprint
  {https://arxiv.org/abs/2012.05357} {arXiv:2012.05357} \BibitemShut {NoStop}%
\bibitem [{\citenamefont {Ominato}\ and\ \citenamefont
  {Koshino}(2012)}]{ominato2012}%
  \BibitemOpen
  \bibfield  {author} {\bibinfo {author} {\bibfnamefont {Y.}~\bibnamefont
  {Ominato}}\ and\ \bibinfo {author} {\bibfnamefont {M.}~\bibnamefont
  {Koshino}},\ }\bibfield  {title} {\bibinfo {title} {Orbital magnetic
  susceptibility of finite-sized graphene},\ }\href
  {https://doi.org/10.1103/PhysRevB.85.165454} {\bibfield  {journal} {\bibinfo
  {journal} {Phys. Rev. B}\ }\textbf {\bibinfo {volume} {85}},\ \bibinfo
  {pages} {165454} (\bibinfo {year} {2012})},\ \Eprint
  {https://arxiv.org/abs/1201.6467} {arXiv:1201.6467} \BibitemShut {NoStop}%
\bibitem [{\citenamefont {Ominato}\ and\ \citenamefont
  {Koshino}(2013)}]{ominato2013}%
  \BibitemOpen
  \bibfield  {author} {\bibinfo {author} {\bibfnamefont {Y.}~\bibnamefont
  {Ominato}}\ and\ \bibinfo {author} {\bibfnamefont {M.}~\bibnamefont
  {Koshino}},\ }\bibfield  {title} {\bibinfo {title} {Orbital magnetism of
  graphene flakes},\ }\href {https://doi.org/10.1103/PhysRevB.87.115433}
  {\bibfield  {journal} {\bibinfo  {journal} {Phys. Rev. B}\ }\textbf {\bibinfo
  {volume} {87}},\ \bibinfo {pages} {115433} (\bibinfo {year} {2013})},\
  \Eprint {https://arxiv.org/abs/1301.5440} {arXiv:1301.5440} \BibitemShut
  {NoStop}%
\bibitem [{\citenamefont {He{\ss}e}\ and\ \citenamefont
  {Richter}(2014)}]{hesse2014}%
  \BibitemOpen
  \bibfield  {author} {\bibinfo {author} {\bibfnamefont {L.}~\bibnamefont
  {He{\ss}e}}\ and\ \bibinfo {author} {\bibfnamefont {K.}~\bibnamefont
  {Richter}},\ }\bibfield  {title} {\bibinfo {title} {Orbital magnetism of
  graphene nanostructures: Bulk and confinement effects},\ }\href
  {https://doi.org/10.1103/PhysRevB.90.205424} {\bibfield  {journal} {\bibinfo
  {journal} {Phys. Rev. B}\ }\textbf {\bibinfo {volume} {90}},\ \bibinfo
  {pages} {205424} (\bibinfo {year} {2014})},\ \Eprint
  {https://arxiv.org/abs/1403.3688} {arXiv:1403.3688} \BibitemShut {NoStop}%
\bibitem [{\citenamefont {Raoux}\ \emph {et~al.}(2014)\citenamefont {Raoux},
  \citenamefont {Morigi}, \citenamefont {Fuchs}, \citenamefont {Pi{\'e}chon},\
  and\ \citenamefont {Montambaux}}]{raoux2014}%
  \BibitemOpen
  \bibfield  {author} {\bibinfo {author} {\bibfnamefont {A.}~\bibnamefont
  {Raoux}}, \bibinfo {author} {\bibfnamefont {M.}~\bibnamefont {Morigi}},
  \bibinfo {author} {\bibfnamefont {J.-N.}\ \bibnamefont {Fuchs}}, \bibinfo
  {author} {\bibfnamefont {F.}~\bibnamefont {Pi{\'e}chon}},\ and\ \bibinfo
  {author} {\bibfnamefont {G.}~\bibnamefont {Montambaux}},\ }\bibfield  {title}
  {\bibinfo {title} {From dia- to paramagnetic orbital susceptibility of
  massless fermions},\ }\href {https://doi.org/10.1103/PhysRevLett.112.026402}
  {\bibfield  {journal} {\bibinfo  {journal} {Phys. Rev. Lett.}\ }\textbf
  {\bibinfo {volume} {112}},\ \bibinfo {pages} {026402} (\bibinfo {year}
  {2014})},\ \Eprint {https://arxiv.org/abs/1306.6824} {arXiv:1306.6824}
  \BibitemShut {NoStop}%
\bibitem [{\citenamefont {Serota}(2000)}]{serota2001}%
  \BibitemOpen
  \bibfield  {author} {\bibinfo {author} {\bibfnamefont {R.~A.}\ \bibnamefont
  {Serota}},\ }\bibfield  {title} {\bibinfo {title} {Van {Vleck} response of a
  two-level system and mesoscopic orbital magnetism of small metals},\ }\href
  {https://doi.org/10.1016/S0038-1098(00)00423-3} {\bibfield  {journal}
  {\bibinfo  {journal} {Solid State Commun.}\ }\textbf {\bibinfo {volume}
  {117}},\ \bibinfo {pages} {99} (\bibinfo {year} {2000})},\ \Eprint
  {https://arxiv.org/abs/cond-mat/0003244} {arXiv:cond-mat/0003244}
  \BibitemShut {NoStop}%
\bibitem [{\citenamefont {Serota}(2002)}]{serota2002}%
  \BibitemOpen
  \bibfield  {author} {\bibinfo {author} {\bibfnamefont {R.~A.}\ \bibnamefont
  {Serota}},\ }\bibfield  {title} {\bibinfo {title} {Mesoscopic fluctuations of
  orbital magnetism in size quantized metals},\ }\href
  {https://doi.org/10.1016/S1386-9477(02)00552-0} {\bibfield  {journal}
  {\bibinfo  {journal} {Physica E}\ }\textbf {\bibinfo {volume} {15}},\
  \bibinfo {pages} {211} (\bibinfo {year} {2002})},\ \Eprint
  {https://arxiv.org/abs/cond-mat/0101456} {arXiv:cond-mat/0101456}
  \BibitemShut {NoStop}%
\bibitem [{\citenamefont {Goldstein}\ and\ \citenamefont
  {Berkovits}(2004)}]{goldstein2004}%
  \BibitemOpen
  \bibfield  {author} {\bibinfo {author} {\bibfnamefont {M.}~\bibnamefont
  {Goldstein}}\ and\ \bibinfo {author} {\bibfnamefont {R.}~\bibnamefont
  {Berkovits}},\ }\bibfield  {title} {\bibinfo {title} {Orbital magnetic
  susceptibility of disordered mesoscopic systems},\ }\href
  {https://doi.org/10.1103/PhysRevB.69.035323} {\bibfield  {journal} {\bibinfo
  {journal} {Phys. Rev. B}\ }\textbf {\bibinfo {volume} {69}},\ \bibinfo
  {pages} {035323} (\bibinfo {year} {2004})},\ \Eprint
  {https://arxiv.org/abs/cond-mat/0308222} {arXiv:cond-mat/0308222}
  \BibitemShut {NoStop}%
\bibitem [{\citenamefont {Lou}\ \emph {et~al.}(2007)\citenamefont {Lou},
  \citenamefont {Wickramasinghe},\ and\ \citenamefont {Serota}}]{lou2007}%
  \BibitemOpen
  \bibfield  {author} {\bibinfo {author} {\bibfnamefont {M.~X.}\ \bibnamefont
  {Lou}}, \bibinfo {author} {\bibfnamefont {J.~M. A. S.~P.}\ \bibnamefont
  {Wickramasinghe}},\ and\ \bibinfo {author} {\bibfnamefont {R.~A.}\
  \bibnamefont {Serota}},\ }\bibfield  {title} {\bibinfo {title} {Mesoscopic
  fluctuations of off-diagonal matrix elements of the angular momentum and
  orbital magnetism of free electrons in a rectangular box},\ }\href
  {https://doi.org/10.1103/PhysRevB.75.224401} {\bibfield  {journal} {\bibinfo
  {journal} {Phys. Rev. B}\ }\textbf {\bibinfo {volume} {75}},\ \bibinfo
  {pages} {224401} (\bibinfo {year} {2007})},\ \Eprint
  {https://arxiv.org/abs/cond-mat/0510156} {arXiv:cond-mat/0510156}
  \BibitemShut {NoStop}%
\bibitem [{\citenamefont {Sticlet}\ and\ \citenamefont
  {Pi{\'e}chon}(2013)}]{sticlet2013}%
  \BibitemOpen
  \bibfield  {author} {\bibinfo {author} {\bibfnamefont {D.}~\bibnamefont
  {Sticlet}}\ and\ \bibinfo {author} {\bibfnamefont {F.}~\bibnamefont
  {Pi{\'e}chon}},\ }\bibfield  {title} {\bibinfo {title} {Distant-neighbor
  hopping in graphene and {Haldane} models},\ }\href
  {https://doi.org/10.1103/PhysRevB.87.115402} {\bibfield  {journal} {\bibinfo
  {journal} {Phys. Rev. B}\ }\textbf {\bibinfo {volume} {87}},\ \bibinfo
  {pages} {115402} (\bibinfo {year} {2013})},\ \Eprint
  {https://arxiv.org/abs/1212.4046} {arXiv:1212.4046} \BibitemShut {NoStop}%
\bibitem [{sup()}]{supplemental}%
  \BibitemOpen
  \href@noop {} {}\bibinfo {note} {See the \hyperref[sm:start]{Supplemental Material} below for derivations,
  the exact crossing configuration, numerical methods, and additional
  calculations.}\BibitemShut {Stop}%
\bibitem [{\citenamefont {Kenyon}(2009)}]{kenyon2009}%
  \BibitemOpen
  \bibfield  {author} {\bibinfo {author} {\bibfnamefont {R.}~\bibnamefont
  {Kenyon}},\ }\bibfield  {title} {\bibinfo {title} {Lectures on dimers},\ }in\
  \href {https://doi.org/10.1090/pcms/016/04} {\emph {\bibinfo {booktitle}
  {Statistical Mechanics}}},\ \bibinfo {series} {IAS/Park City Mathematics
  Series}, Vol.~\bibinfo {volume} {16}\ (\bibinfo  {publisher} {American
  Mathematical Society},\ \bibinfo {address} {Providence, RI},\ \bibinfo {year}
  {2009})\ pp.\ \bibinfo {pages} {191--230},\ \Eprint
  {https://arxiv.org/abs/0910.3129} {arXiv:0910.3129} \BibitemShut {NoStop}%
\bibitem [{\citenamefont {Gade}(1993)}]{gade1993}%
  \BibitemOpen
  \bibfield  {author} {\bibinfo {author} {\bibfnamefont {R.}~\bibnamefont
  {Gade}},\ }\bibfield  {title} {\bibinfo {title} {Anderson localization for
  sublattice models},\ }\href {https://doi.org/10.1016/0550-3213(93)90601-K}
  {\bibfield  {journal} {\bibinfo  {journal} {Nucl. Phys. B}\ }\textbf
  {\bibinfo {volume} {398}},\ \bibinfo {pages} {499} (\bibinfo {year}
  {1993})}\BibitemShut {NoStop}%
\bibitem [{\citenamefont {Motrunich}\ \emph {et~al.}(2002)\citenamefont
  {Motrunich}, \citenamefont {Damle},\ and\ \citenamefont
  {Huse}}]{motrunich2002}%
  \BibitemOpen
  \bibfield  {author} {\bibinfo {author} {\bibfnamefont {O.}~\bibnamefont
  {Motrunich}}, \bibinfo {author} {\bibfnamefont {K.}~\bibnamefont {Damle}},\
  and\ \bibinfo {author} {\bibfnamefont {D.~A.}\ \bibnamefont {Huse}},\
  }\bibfield  {title} {\bibinfo {title} {Particle-hole symmetric localization
  in two dimensions},\ }\href {https://doi.org/10.1103/PhysRevB.65.064206}
  {\bibfield  {journal} {\bibinfo  {journal} {Phys. Rev. B}\ }\textbf {\bibinfo
  {volume} {65}},\ \bibinfo {pages} {064206} (\bibinfo {year} {2002})},\
  \Eprint {https://arxiv.org/abs/cond-mat/0107582} {arXiv:cond-mat/0107582}
  \BibitemShut {NoStop}%
\bibitem [{\citenamefont {Nayak}\ \emph {et~al.}(2024)\citenamefont {Nayak},
  \citenamefont {Sarkar}, \citenamefont {Damle},\ and\ \citenamefont
  {Bera}}]{nayak2024}%
  \BibitemOpen
  \bibfield  {author} {\bibinfo {author} {\bibfnamefont {N.~P.}\ \bibnamefont
  {Nayak}}, \bibinfo {author} {\bibfnamefont {S.}~\bibnamefont {Sarkar}},
  \bibinfo {author} {\bibfnamefont {K.}~\bibnamefont {Damle}},\ and\ \bibinfo
  {author} {\bibfnamefont {S.}~\bibnamefont {Bera}},\ }\bibfield  {title}
  {\bibinfo {title} {Band-center metal-insulator transition in bond-disordered
  graphene},\ }\href {https://doi.org/10.1103/PhysRevB.109.035109} {\bibfield
  {journal} {\bibinfo  {journal} {Phys. Rev. B}\ }\textbf {\bibinfo {volume}
  {109}},\ \bibinfo {pages} {035109} (\bibinfo {year} {2024})},\ \Eprint
  {https://arxiv.org/abs/2304.09674} {arXiv:2304.09674} \BibitemShut {NoStop}%
\bibitem{verbaarschot1994}
J. J. M. Verbaarschot,
Spectrum of the {QCD} {Dirac} operator and chiral random matrix theory,
\href{https://doi.org/10.1103/PhysRevLett.72.2531}{Phys. Rev. Lett. \textbf{72}, 2531 (1994)},
\Eprint {https://arxiv.org/abs/hep-th/9401059} {arXiv:hep-th/9401059}.
\bibitem{altland1997}
A. Altland and M. R. Zirnbauer,
Nonstandard symmetry classes in mesoscopic normal-superconducting hybrid structures,
\href{https://doi.org/10.1103/PhysRevB.55.1142}{Phys. Rev. B \textbf{55}, 1142 (1997)},
\Eprint {https://arxiv.org/abs/cond-mat/9602137} {arXiv:cond-mat/9602137}.
\bibitem [{\citenamefont {Nishigaki}(2012)}]{nishigaki2012b}%
  \BibitemOpen
  \bibfield  {author} {\bibinfo {author} {\bibfnamefont {S.~M.}\ \bibnamefont
  {Nishigaki}},\ }\bibfield  {title} {\bibinfo {title} {Level spacings of
  parametric chiral random matrices and two-color {QCD} with twisted boundary
  condition},\ }\href {https://doi.org/10.1143/PTP.128.1283} {\bibfield
  {journal} {\bibinfo  {journal} {Prog. Theor. Phys.}\ }\textbf {\bibinfo
  {volume} {128}},\ \bibinfo {pages} {1283} (\bibinfo {year} {2012})},\ \Eprint
  {https://arxiv.org/abs/1208.3878} {arXiv:1208.3878} \BibitemShut {NoStop}%
\bibitem [{\citenamefont {Murzaliev}\ \emph {et~al.}(2019)\citenamefont
  {Murzaliev}, \citenamefont {Titov},\ and\ \citenamefont
  {Katsnelson}}]{murzaliev2019}%
  \BibitemOpen
  \bibfield  {author} {\bibinfo {author} {\bibfnamefont {B.}~\bibnamefont
  {Murzaliev}}, \bibinfo {author} {\bibfnamefont {M.}~\bibnamefont {Titov}},\
  and\ \bibinfo {author} {\bibfnamefont {M.~I.}\ \bibnamefont {Katsnelson}},\
  }\bibfield  {title} {\bibinfo {title} {Diamagnetism of metallic nanoparticles
  as a result of strong spin-orbit interaction},\ }\href
  {https://doi.org/10.1103/PhysRevB.100.075426} {\bibfield  {journal} {\bibinfo
   {journal} {Phys. Rev. B}\ }\textbf {\bibinfo {volume} {100}},\ \bibinfo
  {pages} {075426} (\bibinfo {year} {2019})},\ \Eprint
  {https://arxiv.org/abs/1903.05584} {arXiv:1903.05584} \BibitemShut {NoStop}%
\bibitem [{\citenamefont {Forrester}\ and\ \citenamefont
  {Warnaar}(2008)}]{selberg2008}%
  \BibitemOpen
  \bibfield  {author} {\bibinfo {author} {\bibfnamefont {P.~J.}\ \bibnamefont
  {Forrester}}\ and\ \bibinfo {author} {\bibfnamefont {S.~O.}\ \bibnamefont
  {Warnaar}},\ }\bibfield  {title} {\bibinfo {title} {The importance of the
  {Selberg} integral},\ }\href {https://doi.org/10.1090/S0273-0979-08-01221-4}
  {\bibfield  {journal} {\bibinfo  {journal} {Bull. Amer. Math. Soc.}\ }\textbf
  {\bibinfo {volume} {45}},\ \bibinfo {pages} {489} (\bibinfo {year} {2008})},\
  \Eprint {https://arxiv.org/abs/0710.3981} {arXiv:0710.3981} \BibitemShut
  {NoStop}%
\bibitem [{\citenamefont {Wirtz}\ \emph {et~al.}(2015)\citenamefont {Wirtz},
  \citenamefont {Akemann}, \citenamefont {Guhr}, \citenamefont {Kieburg},\ and\
  \citenamefont {Wegner}}]{wirtz2015}%
  \BibitemOpen
  \bibfield  {author} {\bibinfo {author} {\bibfnamefont {T.}~\bibnamefont
  {Wirtz}}, \bibinfo {author} {\bibfnamefont {G.}~\bibnamefont {Akemann}},
  \bibinfo {author} {\bibfnamefont {T.}~\bibnamefont {Guhr}}, \bibinfo {author}
  {\bibfnamefont {M.}~\bibnamefont {Kieburg}},\ and\ \bibinfo {author}
  {\bibfnamefont {R.}~\bibnamefont {Wegner}},\ }\bibfield  {title} {\bibinfo
  {title} {The smallest eigenvalue distribution in the real {Wishart--Laguerre}
  ensemble with even topology},\ }\href
  {https://doi.org/10.1088/1751-8113/48/24/245202} {\bibfield  {journal}
  {\bibinfo  {journal} {J. Phys. A: Math. Theor.}\ }\textbf {\bibinfo {volume}
  {48}},\ \bibinfo {pages} {245202} (\bibinfo {year} {2015})},\ \Eprint
  {https://arxiv.org/abs/1502.03685} {arXiv:1502.03685} \BibitemShut {NoStop}%
\bibitem [{\citenamefont {Jung}\ and\ \citenamefont
  {MacDonald}(2013)}]{jung2013}%
  \BibitemOpen
  \bibfield  {author} {\bibinfo {author} {\bibfnamefont {J.}~\bibnamefont
  {Jung}}\ and\ \bibinfo {author} {\bibfnamefont {A.~H.}\ \bibnamefont
  {MacDonald}},\ }\bibfield  {title} {\bibinfo {title} {Tight-binding model for
  graphene {$\pi$}-bands from maximally localized {Wannier} functions},\ }\href
  {https://doi.org/10.1103/PhysRevB.87.195450} {\bibfield  {journal} {\bibinfo
  {journal} {Phys. Rev. B}\ }\textbf {\bibinfo {volume} {87}},\ \bibinfo
  {pages} {195450} (\bibinfo {year} {2013})}\BibitemShut {NoStop}%
\bibitem{sepioni2010}
M. Sepioni, R. R. Nair, S. Rablen, J. Narayanan, F. Tuna, R. Winpenny,
A. K. Geim, and I. V. Grigorieva,
Limits on intrinsic magnetism in graphene,
\href{https://doi.org/10.1103/PhysRevLett.105.207205}{Phys. Rev. Lett. \textbf{105}, 207205 (2010)},
\Eprint {https://arxiv.org/abs/1007.0423} {arXiv:1007.0423}.
\bibitem{yan2022}
Z. Z. Yan, B. M. Spar, M. L. Prichard, S. Chi, H.-T. Wei,
E. Ibarra-Garc{\'i}a-Padilla, K. R. A. Hazzard, and W. S. Bakr,
Two-dimensional programmable tweezer arrays of fermions,
\href{https://doi.org/10.1103/PhysRevLett.129.123201}{Phys. Rev. Lett. \textbf{129}, 123201 (2022)},
\href{https://arxiv.org/abs/2203.15023}{arXiv:2203.15023}.
\bibitem{mancini2015}
M. Mancini, G. Pagano, G. Cappellini, L. Livi, M. Rider, J. Catani,
C. Sias, P. Zoller, M. Inguscio, M. Dalmonte, and L. Fallani,
Observation of chiral edge states with neutral fermions in synthetic Hall ribbons,
\href{https://doi.org/10.1126/science.aaa8736}{Science \textbf{349}, 1510 (2015)},
\href{https://arxiv.org/abs/1502.02495}{arXiv:1502.02495}.
\bibitem [{\citenamefont {Dorokhov}(1984)}]{dorokhov1984}%
  \BibitemOpen
  \bibfield  {author} {\bibinfo {author} {\bibfnamefont {O.~N.}\ \bibnamefont
  {Dorokhov}},\ }\bibfield  {title} {\bibinfo {title} {On the coexistence of
  localized and extended electronic states in the metallic phase},\ }\href
  {https://doi.org/10.1016/0038-1098(84)90117-0} {\bibfield  {journal}
  {\bibinfo  {journal} {Solid State Commun.}\ }\textbf {\bibinfo {volume}
  {51}},\ \bibinfo {pages} {381} (\bibinfo {year} {1984})}\BibitemShut
  {NoStop}%
\bibitem [{\citenamefont {Kos}\ \emph {et~al.}(2013)\citenamefont {Kos},
  \citenamefont {Nigg},\ and\ \citenamefont {Glazman}}]{kos2013}%
  \BibitemOpen
  \bibfield  {author} {\bibinfo {author} {\bibfnamefont {F.}~\bibnamefont
  {Kos}}, \bibinfo {author} {\bibfnamefont {S.~E.}\ \bibnamefont {Nigg}},\ and\
  \bibinfo {author} {\bibfnamefont {L.~I.}\ \bibnamefont {Glazman}},\
  }\bibfield  {title} {\bibinfo {title} {Frequency-dependent admittance of a
  short superconducting weak link},\ }\href
  {https://doi.org/10.1103/PhysRevB.87.174521} {\bibfield  {journal} {\bibinfo
  {journal} {Phys. Rev. B}\ }\textbf {\bibinfo {volume} {87}},\ \bibinfo
  {pages} {174521} (\bibinfo {year} {2013})},\ \Eprint
  {https://arxiv.org/abs/1303.2918} {arXiv:1303.2918} \BibitemShut {NoStop}%
\end{thebibliography}
\end{document}